\documentclass[twocolumn,aps,pra]{revtex4-1}

\usepackage[usenames,dvipsnames]{xcolor}
\usepackage{amsmath}
\usepackage{amsfonts}
\usepackage{amssymb}
\usepackage[colorlinks=true,citecolor=blue,linkcolor=red]{hyperref}
\usepackage{graphicx}
\usepackage{bbold}					
\usepackage[makeroom]{cancel}		

	\newcounter{subeqn} %
	\makeatletter
	\@addtoreset{subeqn}{equation}
	\makeatother

\DeclareMathOperator{\sign}{sign}  	
\def\abs#1{\left|{#1}\right|}      	
\def\cosrep#1#2{\{{#1}\,|\,\boldsymbol{#2}\}} 
\def\cosrepnb#1#2{\{{#1}\,|\,{#2}\}}
\def\bra#1{\left<{#1}\right|}				
\def\ket#1{\left|{#1}\right>}				
\def\braket#1#2{\left<{#1}|{#2}\right>}		

\def\pder#1#2{\frac{\partial{#1}}{\partial{#2}}}		
\def\bs#1{\boldsymbol{#1}}			

\begin{document}


\title{Weyl semimetal from spontaneous inversion symmetry breaking in pyrochlore oxides}

\author{Tom\'{a}\v{s} Bzdu\v{s}ek}
\author{Andreas R\"{u}egg}
\author{Manfred Sigrist}

\affiliation{Institut f\"{u}r Theoretische Physik, ETH Z\"{u}rich, 8093 Z\"{u}rich, Switzerland}
\date{\today}

\begin{abstract}
We study the electronic properties of strongly spin-orbit coupled electrons on the elastic pyrochlore lattice. Akin to the Peierls transition in one-dimensional systems, the coupling of the lattice to the electronic degrees of freedom can stabilize a spontaneous deformation of the crystal. This deformation corresponds to a breathing mode, which breaks the inversion symmetry. We find that for intermediate values of the staggered strain, the inversion-symmetry broken phase realizes a topological Weyl semimetal. In the temperature-elasticity phase diagram, the Weyl semimetal shows a reentrant phase behavior: it can be reached from a symmetric phase realized both at higher and at lower temperatures. The symmetric phase is a Dirac semimetal, which is protected by the non-symmorphic space group of the pyrochlore lattice. Beyond a critical value of the staggered strain, the symmetry-broken phase is a fully gapped trivial insulator. The surface states of the Weyl semimetal form open Fermi arcs and we observe that their connectivity depends on the termination of the crystal. In particular, for the $\{111\}$ films, the semiclassical closed electronic orbits of the surface states in a magnetic field cross the bulk either twice, four, six or twelve times. We demonstrate how one can tune the number of bulk crossings through a Lifshitz-like transition of the Fermi arcs, which we call Weyl-Lifshitz transition, by applying a surface potential. Our results offer a route to a topological Weyl semimetal in nonmagnetic materials and might be relevant for pyrochlore oxides with heavy transition-metal ions such as alloys of iridates.
\end{abstract}

\maketitle



\section{Introduction}\label{sec:intro}
Pyrochlore oxides experience growing interest as potential hosts of novel electronic phases. Due to the presence of both strong spin-orbit coupling and electronic correlation effects, pyrochlore oxides with heavy transition-metal ions have been identified as a playground for topological phases \cite{Witczak:2014}. Some of the notable theoretical predictions include exotic spin liquids \cite{Machida:2009,Pesin:2010}, topological insulators \cite{Yang:2010,Kargarian:2011,Kurita:2011}, topological crystalline insulators~\cite{Kargarian:2013}, topological semimetals~\cite{Witczak:2012,Witczak:2013} and unusual forms of magnetism~\cite{Gardner:2010,Chen:2012}. Besides these intrinsic three dimensional phases, heterostructuring of pyrochlore oxides offers the possibility to access two-dimensional topological phases, such as the quantum spin Hall or Chern insulator \cite{Hu:2012,Yang:2014,Hu:2014}.

In this work, we add an additional ingredient to the physics of pyrochlore oxides: the coupling of the electronic degrees of freedom to the lattice. Our starting point is a general Hamiltonian on the pyrochlore lattice, which, as a consequence of the non-symmorphic space group, realizes a Dirac semimetal with a discrete set of {\em four-fold} degenerated Fermi points. We find that for an intermediate stiffness of the crystal, an inversion-symmetry breaking staggered strain can spontaneously develop. The symmetry-broken phase realizes a topological Weyl semimetal, whose Fermi surface consists of a discrete set of {\em doubly} degenerated Fermi points with linearly touching valence and conduction bands. If the staggered strain reaches a critical value, the system turns into a fully gapped trivial insulator. Our model shows a rich phase structure including a reentrant Weyl semimetal, which can be reached from the symmetric phase present at both low and high temperatures.

Weyl semimetals have originally been proposed by Wan {\em et. al.}~\cite{Wan:2011} in magnetic pyrochlore iridates, such as Y$_2$Ir$_2$O$_7$. They have been the subject of several reviews \cite{Turner:2013,Hosur:2013, Ramamurthy:2014}. Recent investigations of Weyl semimetals focused on their realizations in interacting models~\cite{Wan:2011,Witczak:2012,Witczak:2013} and in normal insulator-topological insulator multilayers~\cite{Burkov:2011a,Halasz:2012,Zyuzin:2012a}. Recently, Weng {\em et. al.}~\cite{Weng:2015} proposed that certain non-centrosymmetric transition metal monophosphides might be nonmagnetic Weyl semimetals. 

The exciting properties of Weyl semimetals are diverse: First, they host unusual surface states which form disjoint open Fermi arcs rather than closed Fermi lines. The Fermi arcs connect the projections of the Weyl nodes onto the surface Brillouin zone and Fermi arcs on opposite surfaces of the sample are linked through bulk states. These properties lead to characteristic signatures in quantum oscillations experiments in a magnetic field \cite{Potter:2014}. Second, also the bulk properties of Weyl semimetals are unusual due to the chiral anomaly, which states that the electric charge carried by electrons of a {\em given} chirality is \emph{not conserved} in the simultaneous presence of electric and magnetic fields. In contrast, the Nielsen-Ninomiya doubling theorem guarantees that the \emph{net} current carried by electrons of \emph{both} chiralities is conserved~\cite{Nielsen:1981}. The chiral anomaly implies unusual transport properties, reviewed by Hosur and Qi~\cite{Hosur:2013} and by Ramamurthy and Hughes~\cite{Ramamurthy:2014}, which include the axion response~\cite{Chen:2013}, the semiquantized anomalous Hall effect~\cite{Yang:2011,Zyuzin:2012b}, and the (still controversial) chiral magnetic effect~\cite{Chen:2013,Vazifeh:2013,Basar:2014}. A direct way to probe the chiral anomaly in topological semimetals has also been proposed using a non-local transport experiment~\cite{Parameswaran:2014}. 

The Weyl and Dirac semimetals arise naturally as the intermediate phase in a topological insulator - normal insulator phase transition~\cite{Murakami:2007,Murakami:2008,Xu:2011,Yang:2013,Liu:2014c}. More generally, Weyl semimetal can be obtained by starting with a three-dimensional Dirac semimetal and then breaking inversion or time-reversal symmetry~\cite{Burkov:2011a,Burkov:2011b,Halasz:2012,Zyuzin:2012a}. Recent investigation led to the discovery that $\textrm{Na}_{3}\textrm{Bi}$ and $\textrm{Cd}_{3}\textrm{As}_{2}$ are Dirac semimetals~\cite{Wang:2012,Liu:2014a,Liu:2014b,He:2014,Xu:2015}. Young et. al. showed that Dirac semimetal can arise in non-interacting models as a symmetry protected phase~\cite{Young:2012} due to the non-symmorphic elements of the crystal space group. Non-symmorphicity has recently been found to have further effects on the electronic structure, for example, it prohibits band insulators at certain integer fillings~\cite{Parameswaran:2013} or can protect a novel $\mathbb{Z}_2$ topological crystalline insulators~\cite{Liu:2013}.

Our paper is organized as follows. In Sec.~\ref{sec:model}, we introduce a general tight-binding model of spin-orbit coupled electrons on an elastic pyrochlore lattice. In Sec.~\ref{sec:PhaseDiagrams} we discuss how the deformation of the crystal affects the band structure, and we show phase diagrams in the absence and presence of a symmetry-breaking staggered stress field. Due to the importance of the space group in realizing these intriguing electronic phases, we present in Sec.~\ref{sec:GroupTheory} a detailed group theoretical discussion of the spectrum, which particularly highlights the role of non-symmorphicity. In Sec.~\ref{sec:surface} we discuss the surface states of $\{111\}$ and $\{11\bar{1}\}$ oriented films. We show that the surface states depend on the termination of the lattice and introduce the concept of the Weyl-Lifshitz transition, which is characterized by a change in the connectivity of the Fermi arcs.


\section{Model}\label{sec:model}


\subsection{Spin orbit-coupled electrons on the pyrochlore lattice}\label{subsec:SOCmodel}

We study a system of non-interacting electrons on the lattice of corner-sharing tetrahedra called the \emph{pyrochlore} lattice. These tetrahedra form a bipartite lattice and, thus, can be labelled as even and odd [illustrated in bright blue and dark orange in Fig.~\ref{fig:vectors}(d)] in such a way that every even tetrahedron touches only odd ones and vice versa. The pyrochlore lattice is a face-centred cubic (FCC) lattice belonging to a non-symmorphic space group $\# 227$ ($Fd\bar{3}m$). Its Brillouin zone has the form of a truncated octahedron illustrated in Fig.~\ref{fig:vectors}(b).

We consider the following Hamiltonian to describe the dynamics of the electrons on the pyrochlore lattice~\cite{Witczak:2013},
	\begin{eqnarray}
	\mathcal{H}_0 
		&=& \sum_{\left<i,j\right>} c_i^\dagger\left( t_1 + it_2 \boldsymbol{d}_{ij}\cdot\boldsymbol{\tau}\right)c_j^{\phantom{\dagger}} + \nonumber\\
		&\phantom{=}& + \sum_{\left<\left<i,j\right>\right>} c_i^\dagger\left[t_1' + i\left(t_2'\boldsymbol{\mathcal{R}}_{ij} + t_3'\boldsymbol{\mathcal{D}}_{ij}\right)\cdot\boldsymbol{\tau}\right]c_j^{\phantom{\dagger}}. \label{eqn:H0}
	\end{eqnarray}
The electron annihilation operator at site $i$ is given by $c_i=(c_{i\uparrow},c_{i\downarrow})^T$, where $\alpha=\uparrow,\downarrow$ refers to a (pseudo-)spin-1/2 degree of freedom. The meaning of the symbols is as follows: The first sum runs over the nearest neighbor (NN) and the second sum over the next-nearest neighbor (NNN) bonds. The Pauli matrices $\boldsymbol{\tau} = (\tau^x,\tau^y,\tau^z)$ describe a global basis in spin space, and the terms containing them represent spin-orbit coupling. The real space vectors appearing in the NN part of $\mathcal{H}_0$ are defined via 
	\begin{subequations}
	\begin{align}
	\boldsymbol{d}_{ij} &= 2\boldsymbol{a}_{ij}\times\boldsymbol{b}_{ij}\\
	\boldsymbol{a}_{ij} &= \tfrac{1}{2}\left(\boldsymbol{b}_i + \boldsymbol{b}_j\right) -\boldsymbol{x}_\textrm{c}\\
	\boldsymbol{b}_{ij} &= \boldsymbol{b}_j - \boldsymbol{b}_i \refstepcounter{equation} \\
	\boldsymbol{x}_\textrm{c} & = \tfrac{1}{2}\left(\boldsymbol{b}_1 + \boldsymbol{b}_2 + \boldsymbol{b}_3 + \boldsymbol{b}_4\right)	
	\end{align}
	\end{subequations}	
and those in the NNN part are defined as
	\begin{subequations}\label{eqn:sub2}
	\begin{align}
	\boldsymbol{\mathcal{R}}_{ij} &= \boldsymbol{b}_{ik}\times\boldsymbol{b}_{kj} \refstepcounter{equation}\label{eqn:s1} \\
	\boldsymbol{\mathcal{D}}_{ij} &= \boldsymbol{d}_{ik}\times\boldsymbol{d}_{kj}  
	\end{align}
	\end{subequations}
where $k$ is the common NN of the NNN sites $i$ and $j$. We further use position vectors $\boldsymbol{b}_i$ pointing to the site $i$ of the tetrahedron 
	\begin{subequations}\label{eqn:subB}
	\begin{align}
	\bs{b}_1 &= a(0,0,0) \\
	\bs{b}_2 &= a(0,1,1) \\
	\bs{b}_3 &= a(1,0,1) \\
	\bs{b}_4 &= a(1,1,0) 
	\end{align}
	\end{subequations}
and position vector $\boldsymbol{x}_\textrm{c}$ pointing to the centre of the tetrahedron
	\begin{equation}
	\bs{x}_{\textrm{c}} = \frac{1}{4}\sum_{i=1}^4 \bs{b}_i
	\end{equation}
Examples of $\boldsymbol{b}_{ij}$, $\boldsymbol{a}_{ij}$ and $\boldsymbol{d}_{ij}$ are illustrated in Fig.~{\ref{fig:vectors}(a)}. We denote the length of the cube circumscribed to the tetrahedra [Fig.~{\ref{fig:vectors}(a)] as $a$ so that $\abs{\boldsymbol{b}_{ij}}=a\sqrt{2}$. The edge length of the FCC unit cell is $4a$.

Hamiltonian (\ref{eqn:H0}) is the most general single-orbital Hamiltonian with spin-orbit coupling and up to NNN terms respecting the full symmetry of the pyrochlore lattice ~\cite{Kurita:2011}. Such a situation arises for the five $5d$-electrons coming from each $\textrm{Ir}^{4+}$ site of the pyrochlore oxides. The crystal field of the neighbouring atoms splits the $5d$ orbitals into the six $t_{2g}$ states and the four $e_g$ states, the latter being higher in energy. The on-site spin-orbit coupling further splits the six degenerate $t_{2g}$ states into an effective pseudospin $j_\textrm{eff} = 1/2$ doublet and an effective $j_\textrm{eff}=3/2$ quadruplet, the latter being lower in energy. The five $5d$-electrons of $\textrm{Ir}^{4+}$ sites fill the $j_\textrm{eff}=3/2$ quadruplet and half-fill the $j_\textrm{eff} = 1/2$ doublet~\cite{Pesin:2010}. Hamiltonian (\ref{eqn:H0}) can then be viewed as an effective model for electrons residing in the $j_\textrm{eff} = 1/2$ orbitals. In this way one can relate the parameters $t_{1,2}$ and $t_{1,2,3}'$ to particular orbital overlaps as is thoroughly explained in Ref.~\cite{Witczak:2013}. 
\begin{figure}[t!]
	\includegraphics[width=0.48\textwidth]{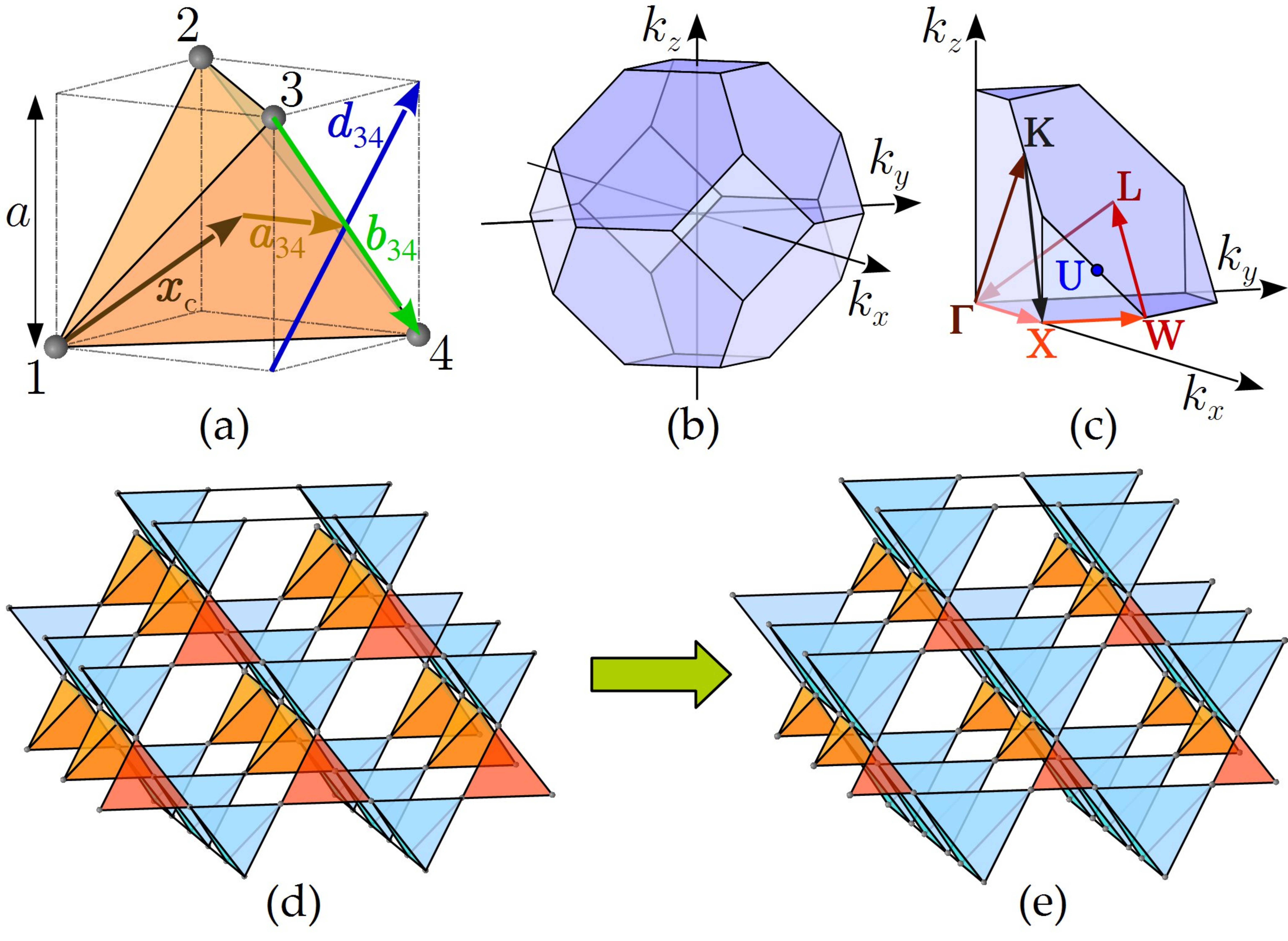}  
  \caption{(Color online) (a) Illustration of vectors $\boldsymbol{x}_{\textrm{c}}$, $\boldsymbol{a}_{ij}$, $\boldsymbol{b}_{ij}$, and $\boldsymbol{d}_{ij}$ used in Hamiltonians (\ref{eqn:H0}) and (\ref{eqn:Hdelta}). (b) Shape of the Brillouin zone (BZ) of a face-centred cubic (FCC) lattice. This corresponds to both the pyrochlore lattice and the breathing pyrochlore lattice. (c) Definition of the high-symmetry points of the FCC lattice within one eighth of the Brillouin zone and the path $\Gamma\textrm{XWL}\Gamma\textrm{KX}$ in the momentum space used for plotting energy spectra in Fig.~\ref{fig:spectra}. The $\textrm{U}$ point is equivalent to the $\textrm{K}$ point. (d) Pyrochlore lattice with differently colored even (blue/bright) and odd (orange/dark) tetrahedra. (e) The $\mathcal{I}$-broken (breathing) pyrochlore lattice considered in the elastic model. 
  }
  \label{fig:vectors}
\end{figure}

A generic spectrum of Hamiltonian (\ref{eqn:H0}) is plotted in Fig.~\ref{fig:spectra}(a). Note that the site fillings $n=1/4$ and $n=3/4$ correspond to a semimetallic phase. 
In pyrochlore iridates, such a commensurate filling may be realized by considering alloys of the form A$_{2-x}$B$_x$Ir$_2$O$_7$ where A and B are nonmagnetic but have different oxidation states \cite{Fukazawa:2002,Soda:2003,CosioCastaneda:2013}. For example, we expect that Y$_{1.5}$Ca$_{0.5}$Ir$_2$O$_7$ \cite{Fukazawa:2002} realizes the site filling $n=1/4$ and the (hypothetical) compound Bi$_{1.5}$Se$_{0.5}$Ir$_2$O$_7$ site filling $n=3/4$ in our model. To be concrete,
we will consider 
	\begin{equation} 
	n=3/4\label{eqn:filling}
	\end{equation} 
throughout the paper.


\subsection{Elastic lattice}\label{subsec:elastic}

Electron-phonon coupling can lead to softening of certain phonon modes and to distortion of the lattice akin to the Peierls transition. This is also similar to certain valence bond solids arising due to interactions~\cite{Indergand:2006}. The leading instabilities can be found by investigating the Lindhard function. We argue that the leading lattice instability of model (\ref{eqn:H0}) occurs at momentum $\bs{q}=\bs{0}$. To illustrate this, we ignore the two occupied bands far from the chemical potential and consider only the conduction and the valence bands. Defining $\xi_{\alpha,\bs{k}}=\varepsilon_{\alpha,\bs{k}}-\mu$ where $\alpha$ stands for the band index, Fig.~\ref{fig:spectra}(a) indicates approximate electron-hole symmetry 
	\begin{equation}
	(\xi_{\textrm{con},\bs{k}}-\mu)\approx -(\xi_{\textrm{val},\bs{k}}-\mu)
	\end{equation}
In the case of a \emph{perfect} electron-hole symmetry $\xi_{\textrm{con},\bs{k}}=-\xi_{\textrm{val},\bs{k}}$, the static ($\omega=0$) Lindhard function at zero temperature satisfies
	\begin{eqnarray}
	0<-\chi{\left(\bs{q}\right)}
		&=&-\frac{1}{\Omega}\sum_{\bs{k}\in\textrm{BZ}}\sum_{\alpha,\alpha'}\frac{f(\xi_{\alpha,\bs{k}-\bs{q}})-f(\xi_{\alpha',\bs{k}})}{\xi_{\alpha,\bs{k}-\bs{q}}-\xi_{\alpha',\bs{k}}} \nonumber\\ 
	&=& \frac{1}{\Omega} \sum_{\bs{k}\in\textrm{BZ}}\frac{2}{\xi_{\bs{k}-\bs{q}}+\xi_{\bs{k}}}\nonumber\\
	&\leq& \frac{1}{\Omega} \sum_{\bs{k}\in\textrm{BZ}} \left(\frac{1}{2\xi_{\bs{k}-\bs{q}}}+\frac{1}{2\xi_{\bs{k}}}\right) = - \chi{\left(\bs{0}\right)}
	\end{eqnarray}
where we used the arithmetic-harmonic mean inequality. The \emph{equality sign} applies only if $\bs{q}=\bs{0}$ or if both bands are perfectly flat. The result means that the Lindhard function has a peak at $\bs{q}=\bs{0}$ which corresponds to the leading instability. We expect this peak to be preserved for Hamiltonian (\ref{eqn:H0}) where the electron-hole symmetry is approximately valid.

Inspired by this observation, we consider the \emph{most symmetric} $\bs{q}=\bs{0}$ phonon mode which corresponds to the simultaneous expansion of the even and shrinking of the odd tetrahedra as illustrated in Fig.~\ref{fig:vectors}(e). We call it the \emph{breathing mode} of the pyrochlore lattice. Similar breathing pyrochlore lattice has been observed in certain $A$-site ordered spinel oxides~\cite{Okamoto:2013,Kimura:2014}. 

We treat the breathing mode classically and refer to its amplitude as \emph{staggered strain} $\delta$. We model its effect on the electron Hamiltonian by multiplying the NN terms by a factor $(1+(-)\delta)$ for the short (long) bonds. We ignore the higher order influence on both the NN and the NNN terms as these are assumed to have a quantitative but not qualitative effect on the phase diagram of the model. The elastic energy of the lattice deformation is set to be proportional to the square of the amplitude $\delta$. The complete Hamiltonian then reads
	\begin{eqnarray}
	\mathcal{H}_\delta 
		&=&  \frac{1}{2} Y\Omega \delta^2 + \sum_{\left<i,j\right>} \left(1\pm\delta\right)c_i^\dagger\left(t_1  + it_2 \boldsymbol{d}_{ij}\cdot\boldsymbol{\tau}\right)c_j^{\phantom{\dagger}} \nonumber\\
		&\phantom{=}& + \sum_{\left<\left<i,j\right>\right>} c_i^\dagger\left[t_1' + i\left(t_2'\boldsymbol{\mathcal{R}}_{ij} + t_3'\boldsymbol{\mathcal{D}}_{ij}\right)\cdot\boldsymbol{\tau}\right]c_j^{\phantom{\dagger}}  \label{eqn:Hdelta}
	\end{eqnarray}
where $Y$ is elasticity of the lattice analogous to the Young modulus, and $\Omega$ is the volume of the sample. The equilibrium value of $\delta$ is determined by minimizing the energy of the electron-lattice system which, by the Hellmann--Feynman theorem, corresponds to solving the self-consistency equation
	\begin{equation}
	\delta = \frac{1}{Y\Omega}\sum_{\left<i,j\right>}\bra{\Psi_\delta}\mp c_i^\dagger\left(t_1  + it_2 \boldsymbol{d}_{ij}\cdot\boldsymbol{\tau}\right)c_j^{\phantom{\dagger}}\ket{\Psi_\delta}\label{eqn:self-con-for-delta}
	\end{equation}
where $\ket{\Psi_\delta}$ is the ground state of the electron Hamiltonian for a staggered strain amplitude $\delta$. 

Hamiltonians (\ref{eqn:H0}) and (\ref{eqn:Hdelta}) have many free parameters. To simplify the situation, we reduce the parameter space by setting 
	\begin{subequations}
	\begin{equation}
	R:=\frac{t_2}{t_1}=\frac{t_2'}{t_1'}=\frac{t_3'}{t_1'}
	\end{equation}
for the relative strength of the spin orbit coupling and 
	\begin{equation}
	p:= \frac{t_1'}{t_1}=\frac{t_2'}{t_2}
	\end{equation}  
for the relative strength of the NNN terms. We further define a variable 
	\begin{equation}
	s=\sign(t_1)=\pm 1.
	\end{equation}
	\end{subequations}
These substitutions modify Hamiltonian (\ref{eqn:Hdelta}) to
\begin{eqnarray}
\mathcal{H}_\delta 
	&=&  \frac{1}{2} Y\Omega \delta^2 + s\Big\{\sum_{\left<i,j\right>} \left(1\pm\delta\right)c_i^\dagger\left(1  + iR \boldsymbol{d}_{ij}\cdot\boldsymbol{\tau}\right)c_j^{\phantom{\dagger}}  \nonumber \\
	&\phantom{=}& \quad + \, p \sum_{\left<\left<i,j\right>\right>} c_i^\dagger\left[1 + i R\left(\boldsymbol{\mathcal{R}}_{ij} + \boldsymbol{\mathcal{D}}_{ij}\right)\cdot\boldsymbol{\tau}\right]c_j^{\phantom{\dagger}}\Big\}  \label{eqn:Hsimp}
\end{eqnarray} 
where all energies are now in units of $\abs{t_1}$. We will work with 
	\begin{subequations}\label{eqn:subChosenParam}
	\begin{align}
	R&=-0.4 \\
	p&=-0.1 \\ 
	s&=+1 
	\end{align}
	\end{subequations}
unless otherwise stated. This choice approximately corresponds to the parameters used in Ref.~\cite{Witczak:2013}.

It may happen that the energy of the Hamiltonian (\ref{eqn:Hsimp}) is minimized for a non-zero value of $\delta$. This means that the cost of the deformation is compensated for by the electron energies, thus making the lattice deformation energetically favorable. Such a transition decreases the symmetry of the lattice. In particular, it breaks the inversion symmetry $\mathcal{I}$. The $\mathcal{I}$-broken lattice still has the FCC Bravais lattice and an unchanged Brillouin zone but it belongs to a symmorphic $\# 216$ ($F\bar{4}3m$) space group.



\section{Phase diagrams}
\label{sec:PhaseDiagrams}
\subsection{Evolution of the band structure}\label{subsec:bands}

We start by describing the evolution of the spectrum of Hamiltonian (\ref{eqn:Hsimp}) as we tune the staggered strain amplitude $\delta$. The following observations are based on a direct numerical diagonalization of the Hamiltonian and are summarized in Fig.~\ref{fig:spectra}. In Sec.~\ref{sec:GroupTheory}, we will present detailed group theoretical arguments.
\begin{figure}[hbt!]
\centering
\includegraphics[width=0.48\textwidth]{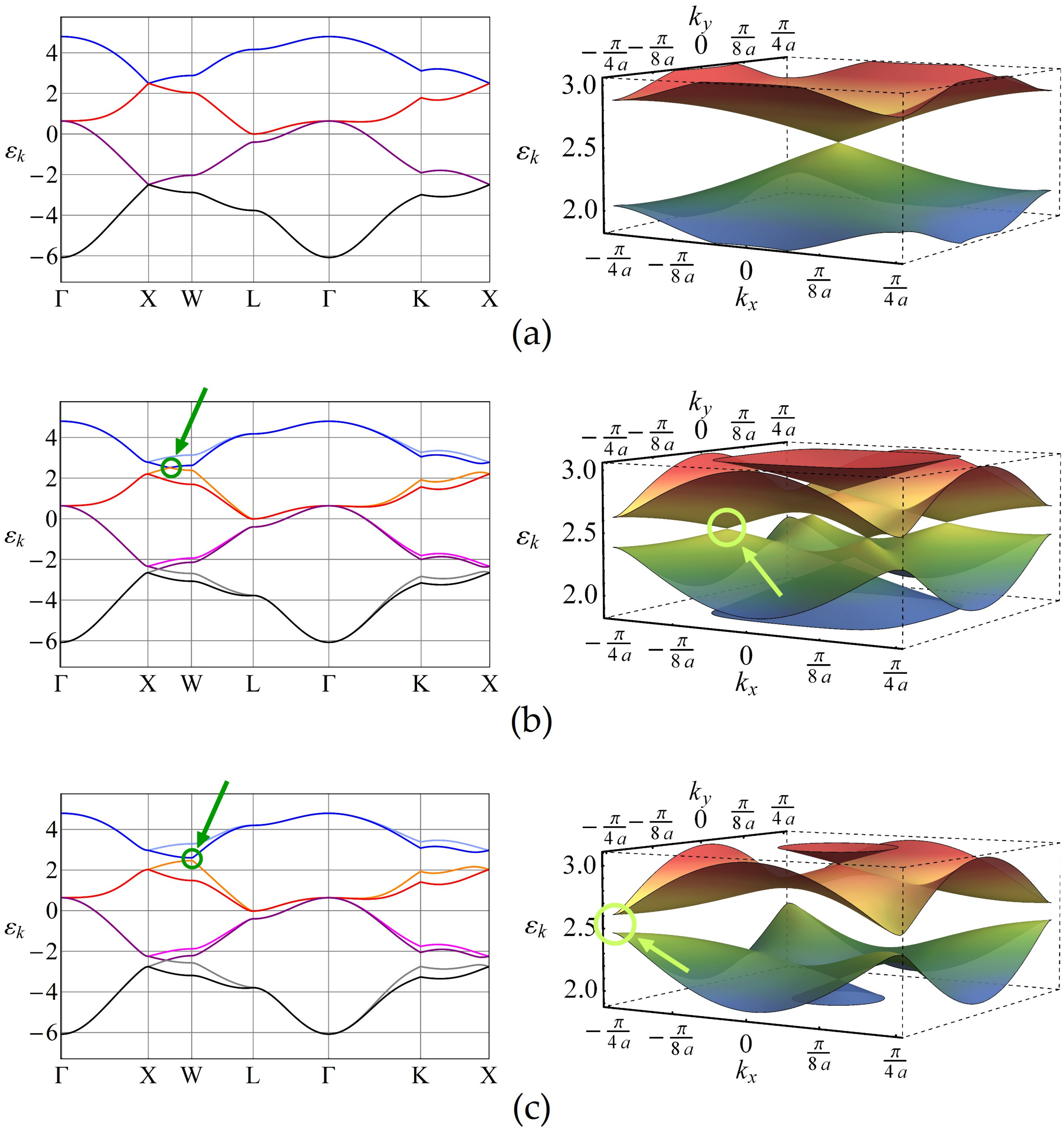} 
\caption{(Color online) The spectrum of Hamiltonian (\ref{eqn:Hsimp}) for parameters (\ref{eqn:subChosenParam}) and (a) $\delta = 0$, which corresponds to the Dirac semimetal (DSM), (b) $\delta = 0.08$, which corresponds to the Weyl semimetal (WSM) with 12 distinct Weyl nodes, and (c) $\delta=0.13$, which corresponds to a trivial band insulator (INS). Left figures show the spectra in the three cases along the path in the BZ indicated in Fig.~\ref{fig:vectors}(c), right figures show the spectra over the square face of the BZ. The arrows indicate position of one of the Weyl nodes for the WSM phase, and the place where the Weyl nodes have annihilated for the INS phase.}
\label{fig:spectra}
\end{figure}

In the absence of a staggered strain, $\delta=0$ [Fig.~\ref{fig:spectra}(a)], we find that all bands are doubly degenerate as a consequence of the simultaneous presence of time-reversal $\mathcal{T}$ and inversion symmetry $\mathcal{I}$. At each $\textrm{X}$ point [see Fig.~\ref{fig:vectors}(c) for the definition of the high-symmetry points], we find two energetically separated Dirac nodes where four bands reach the same energy and disperse linearly in all directions. This is similar to the spectrum of graphene but in three rather than just two dimensions. Since the inequivalent $\textrm{X}$ points are related by crystal symmetries, all higher (and all lower) Dirac nodes are realized at the same energy. It is therefore possible to tune the chemical potential to this value, which corresponds to a site filling of $n=3/4$ (six electrons per unit cell) for the upper and $n=1/4$ (two electrons per unit cell) for the lower Dirac nodes. Such a system has a Fermi surface consisting of a set of $\bs{k}$-points and is usually referred to as a \emph{Dirac semimetal} (DSM)~\cite{Murakami:2007,Young:2012,Wang:2012,Liu:2014a,Liu:2014b,Xu:2015,He:2014}. 

A non-vanishing staggered strain, $\delta\neq 0$ [Fig.~\ref{fig:spectra}(b)], leads to a splitting of each Dirac node into four \emph{Weyl nodes}. These are points where only two rather than four bands touch each other, and they disperse linearly in all directions around this point. As we increase the value of $\delta$, the Weyl nodes move along the $\textrm{XW}$ lines. The  upper 12 Weyl nodes of the model are mutually related by crystal symmetries and time-reversal. As a consequence, they are all realized at the same energy and the chemical potential resumes to be tuned to them if either $n=1/4$ or $n=3/4$. Such a phase is called the \emph{Weyl semimetal} (WSM)~\cite{Wan:2011,Hosur:2013,Turner:2013,Ramamurthy:2014}. 
\begin{figure}[t!]
  \begin{center}
	\includegraphics[width=0.33\textwidth]{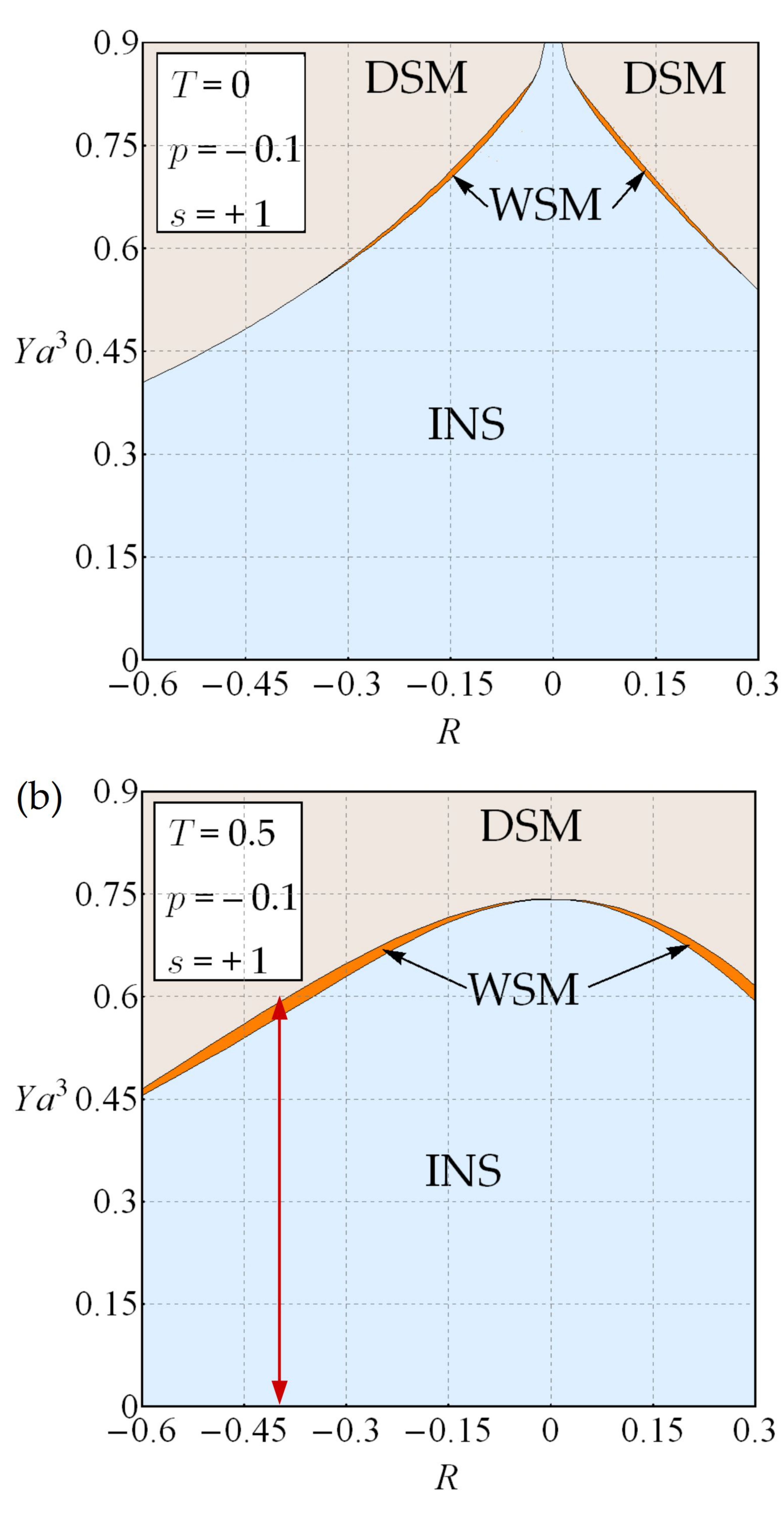}
	\end{center}
	\caption{(Color online) Phase diagrams of Hamiltonian (\ref{eqn:Hsimp}) for $p=-0.1$ and $s=+1$ in the $R-Y$ plane for (a) $T=0$, and (b) $T=0.5$. The red (vertical) arrow in diagram (b) corresponds to the same set of parameters as the (horizontal) red arrow in Fig.~\ref{fig:wings-all}(a). At $T=0$ ($T=0.5$), the transition from the DSM phase to the WSM phase is first (second) order for the shown parameters.}
  \label{fig:phase-diagrams}
\end{figure}

In the vicinity of each Weyl node, we can approximate the electron Hamiltonian as
\begin{equation}
\mathcal{H}(\bs{k}) = (E_0 + u_i k_i)\mathbb{1} + v_{ij}k_i\sigma_j \label{eqn:WeylHam}
\end{equation}
where $\boldsymbol{\sigma}=(\sigma_x,\sigma_y,\sigma_z)$ are the Pauli matrices, $\det[v_{ij}]\neq 0$ and vector $\bs{u}$ describes the tilt of the dispersion cone. The quantity $\sign(\det[v])=\pm 1$ associated with each Weyl node is called \emph{chirality}. It gives rise to the topological nature of the WSM, which manifests itself, e.g., by the appearance of the Fermi arcs in the surface Brillouine zone~\cite{Wan:2011} and by the chiral anomaly~\cite{Hosur:2013}. The robustness of the WSM against all local perturbations follows from the fact that all Pauli matrices are used in the effective low energy Hamiltonian (\ref{eqn:WeylHam}) \cite{Wan:2011}. Gapping the spectrum requires enlarging of the effective low-energy Hilbert space and can be achieved either by scattering between different Weyl nodes (which breaks translational symmetry) or by forming a superconducting state (which requires breaking the global $U(1)$ symmetry)~\cite{Hosur:2013}.

Time-reversal symmetry implies that if there is a Weyl node at momentum $\bs{k}$, then there is a Weyl node of the \emph{same} chirality at $-\bs{k}$. On the other hand, Weyl points at momenta related by mirror symmetry carry opposite chiralities. These requirements allow us to characterize all twelve Weyl nodes. We find 6 Weyl nodes with positive and six with negative chirality, in accordance with the Nielsen-Ninomiya doubling theorem~\cite{Nielsen:1981}.

At a critical value of the staggered strain $\delta=\delta_\textrm{c}$, pairs of Weyl nodes with opposite chirality meet at the $\textrm{W}$ points where they annihilate. The spectrum disperses quadratically along the $\textrm{XW}$ lines, which correspond to the direction of motion of the Weyl nodes~\cite{Murakami:2008}. Such a quadratic band touching point does not carry chirality~\cite{Liu:2014c}. Finally, for $\delta > \delta_\textrm{c}$, we find a gapped phase, which is a topologically trivial band insulator (INS). For the chosen parameters (\ref{eqn:subChosenParam}) and filling factor~(\ref{eqn:filling}), the critical staggered strain is 
	\begin{equation}
	\delta_\textrm{c}\approx 0.1112.\label{eqn:crit_delta}
	\end{equation}
%


\subsection{Phase diagrams for elastic lattices}
\label{subsec:phases}
To find the phase diagrams at filling $n=3/4$, we numerically solve the self-consistency equation (\ref{eqn:self-con-for-delta}) for varying hopping parameters and elasticities of the lattice. In general, we expect that the deformation of a stiff lattice (i.e.~with a large value of $Y$) is energetically too costly, so that the system will remain in the $\mathcal{I}$-preserving DSM phase. Decreasing the elasticity should allow for a transition to an $\mathcal{I}$-broken state that can be either a WSM or an INS. 

We first study the role of the relative spin-orbit coupling $R$, see Figs.~\ref{fig:phase-diagrams}(a) and (b). Throughout this subsection, we fix $p=-0.1$ and $s=+1$ and we vary $R$ in the range $(-0.774,+0.364)$ for which the undistorted system realizes a DSM phase. For values of $R$ outside of this interval, the valence bands at $\Gamma$ rise above the Dirac nodes at $\textrm{X}$, leading to a metallic state with a hole-like Fermi pocket at the $\Gamma$ point and three electron-like Fermi pockets at the inequivalent $\textrm{X}$ points. The value $R=0$ corresponds to no spin-orbit coupling and hence spin-independent hopping amplitudes. In this case, the spectrum does not exhibit Dirac nodes but line nodes along the $\textrm{XW}$ lines. This line degeneracy is gapped out by any $\delta\neq 0$. The transition from $R<0$ to $R>0$ also changes the degeneracies at the $\Gamma$ point (bottom-to-top) from $2-4-2$ to $4-2-2$.
\begin{figure*}[t!]
  \begin{center}
	\includegraphics[width=0.995\textwidth]{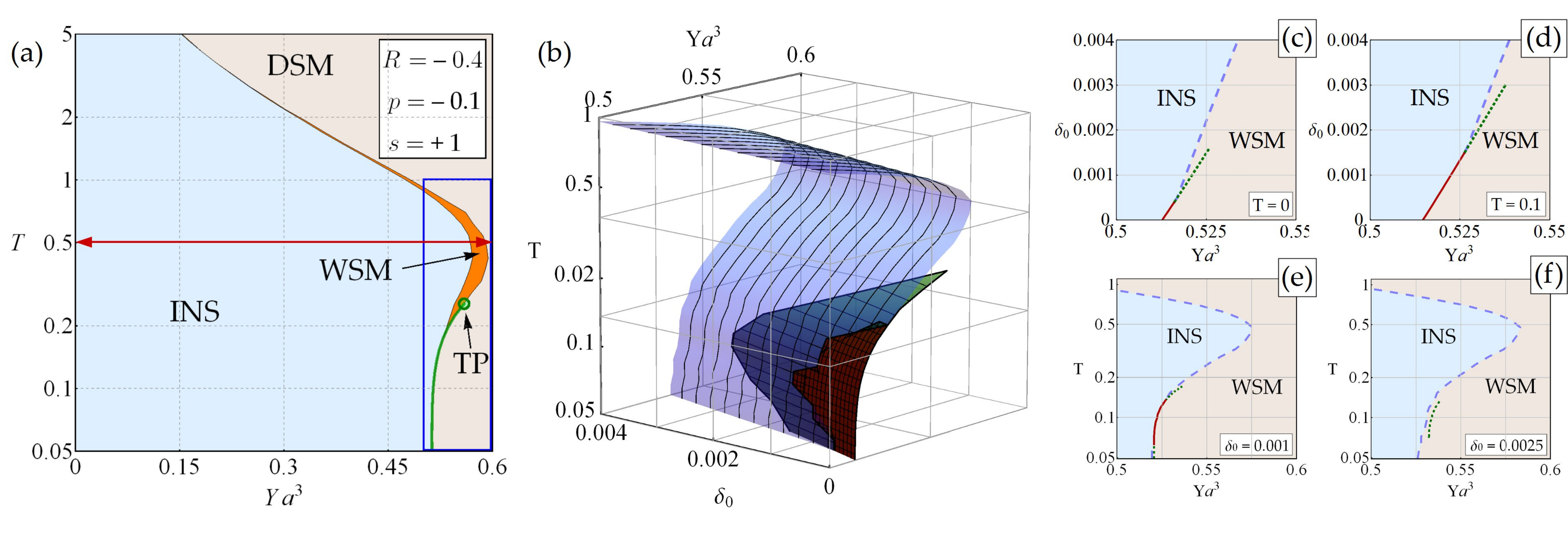} 
  \end{center}
  \caption{(Color online) (a) Phase diagram of Hamiltonian (\ref{eqn:Hsimp}) for $R=-0.4$, $p=-0.1$, $s=+1$ in the $R-T$ plane. The WSM phase is most robust at intermediate temperatures. The thick (green) line indicates the first-order transition between the symmetric DSM and the symmetry-broken WSM and INS phases and ends at a tricritical point TP. At higher temperatures, the transition is always second order. The horizontal (red) arrow corresponds to the same set of parameters as the vertical (red) arrow in Fig.~\ref{fig:phase-diagrams}(b). The blue rectangle corresponds to the range of temperatures and elasticities that are used in the three-dimensional diagram. (b) Phase diagram for the same parameters with included coupling to a symmetry-breaking staggered strain $Y\delta_0$. In the presence of such a field, the distinction between the DSM and the WSM phases ceases to have a meaning. Only the structural first-order transition (the dark green sheet in the back) and the transition between the WSM and INS phases (the pale blue sheet in the front) are therefore present. Where the two transitions occur at the same parameters, the dark red sheet is plotted. (c,d) Horizontal cuts of the three-dimensional phase diagram in (b) corresponding to fixed temperature, and (e,f) vertical cuts corresponding to a fixed staggered strain. In these four diagrams, the dotted green lines correspond to the first-order structural transition, the dashed blue lines to the WSM-to-INS transition, and the solid red lines is drawn where the two transitions coincide.}
  \label{fig:wings-all}
\end{figure*}

At zero temperatures, see Fig.~\ref{fig:phase-diagrams}(a), the transition from $\delta = 0$ to $\delta \neq 0$ is always first order. Furthermore, if $R\in (-0.35,0.28)\backslash\{0\}$, a narrow WSM phase is found by varying the elasticity. (Our numerical accuracy does not allow a definite conclusion on the presence of the WSM phase for $\abs{R}\lesssim 0.02$). For values of $R$ outside of the mentioned interval, we find a direct first-order transition between the DSM and the INS phase, i.e., the staggered strain directly jumps to a value $\delta>\delta_c$.

In an equivalent calculation at $T=0.5$, the transition is found to be second order for all values of $R$. By reducing $Y$, $\delta$ smoothly increases from $0$ in the DSM to a value $\delta>\delta_\textrm{c}$ in the INS, leading to a larger WSM region than at zero temperature.

We further study the phase diagram in the $Y$-$T$ plane for fixed $R=-0.4$ [Fig.~\ref{fig:wings-all}(a)]. The diagram demonstrates that the WSM phase is indeed most robust at intermediate temperatures. We further observe a reentrant phase behavior: Starting in the symmetric Dirac semimetal at high temperatures, the system spontaneously breaks the inversion symmetry upon cooling. But upon further cooling, it returns to the symmetric phase. For the chosen parameters, we observe this behavior in the range $Ya^3\in (0.513,0.594)$. The reentrance can be traced back to the peculiar form of the density-of-states (DOS) in the symmetric DSM, which has a local minimum with a vanishing DOS at the Fermi energy. Thermal broadening then {\em enhances} the effective DOS at the chemical potential, rendering the system more susceptible to a symmetry-breaking transition at elevated temperatures than at low temperatures. A similar reentrant phase behavior has also been observed in models for metallic metamagnetic systems \cite{Binz:2004}.

At low temperatures, the transition between the symmetric and the symmetry-broken phase is first order. The first-order transition line obeys the Clausius-Clapeyron relation, and has to approach zero temperature perpendicularly. The first-order transition might also lead to a hysteretic behavior with temperature. Slightly below the tip of the belly-shaped WSM phase, there is a tricritical point (TP). For temperatures above the tricritical point, the transition between the symmetric and the symmetry-broken phase is second order.


\subsection{Phase diagram in a symmetry-breaking field}
\label{subsec:wings}

To further explore the interesting structure of the phase diagram, we study the effect of a symmetry-breaking field, which is conjugate to the staggered strain (i.e.~a staggered stress). We incorporate the symmetry-breaking field in the elastic lattice model (\ref{eqn:Hsimp}) by modifying the elastic energy term as follows,
	\begin{equation}
	\frac{1}{2}Y\Omega \delta^2 \mapsto \frac{1}{2}Y\Omega \left(\delta-\delta_0\right)^2 \label{eqn:sym-break-field},
	\end{equation}
where $Y\delta_0$ parametrizes the staggered stress. The resulting three-dimensional phase diagram in the $(Y,\delta_0,T)$-space is shown in Fig.~\ref{fig:wings-all}(b). Due to the complexity of the three-dimensional diagram in Fig.~\ref{fig:wings-all}(b), we show in Figs.~\ref{fig:wings-all}(c)-(f) several slices at either a fixed temperature or a fixed staggered stress. 

A non-vanishing $\delta_0$ explicitly breaks the inversion symmetry. Therefore, the DSM changes into a WSM phase with a small separation of the Weyl nodes proportional to $\delta_0$. This means that the \emph{second-order} transition line separating DSM from WSM at $\delta_0=0$ ceases to exist in the presence of a staggered stress. However, the first-order line below the triciritical point survives also for $\delta_0>0$, forming a sheet of first-order transitions (a so-called \emph{Griffiths wing}), which extends up to a finite value of $\delta_0$. The Griffiths wing either signals a structural transition \emph{within} the WSM phase [dark green sheet in Fig.~\ref{fig:wings-all}(b) in the online version] or {\em between} the WSM and the INS phases [dark red sheet in Fig.~\ref{fig:wings-all}(b) in the online version]. Note that the Griffiths wing takes an unusual form with a ``belly" at finite temperatures: the end point at $T=0.1$ occurs at $\delta_0\approx 0.0030$ while at $T=0$ it occurs at $\delta_0\approx 0.0016$, see Figs.~\ref{fig:wings-all}(c) and (d). Finally, we note that the boundary between the WSM phase and the INS phase remains well-defined for all values of $\delta_0$ and $T$ [light blue sheet in Fig.~\ref{fig:wings-all}(b) in the online version].


\section{Group theoretical analysis}
\label{sec:GroupTheory}


\subsection{The general strategy}

In this section, we will demonstrate how the symmetry of the pyrochlore lattice inevitably leads to the Dirac node at the $\mathrm{X}$ point of the Brillouin zone, and why this Dirac node has to split into four Weyl nodes upon breaking the inversion symmetry. The pyrochlore lattice belongs to the same space group as the diamond lattice ($\#227$, $Fd\bar{3}m$) so the same reasoning as that of Ref.~\cite{Young:2012} applies. Further structures belonging to this space group are $\beta$-cristobalite and spinel oxides. 

Our main tool in determining the spectrum degeneracies at a given $\bs{k}$-point is the following relation between representations of point-symmetry operations in the vector space spanned by the Bloch wave functions at $\bs{k}$:
	\begin{equation}
	\overline{D}_{\bs{k}}(R_i)\overline{D}_{\bs{k}}(R_j)= \exp{(-i\bs{g}_i\cdot\bs{t}_j)} \overline{D}_{\bs{k}}(R_i \circ R_j).\label{eqn:projrec}
	\end{equation}
Here, $R_i$ is a point-symmetry operation that maps the considered $\bs{k}$-point onto itself modulo a reciprocal lattice vector $\bs{g}_i=(R_i^{-1}\bs{k})-\bs{k}$, and $\bs{t}_i$ is the non-symmorphic shift associated with the point operation $R_i$. The set of all such $\bs{k}$-preserving operations $R_i$ forms a group $\overline{G}^{\bs{k}}$ called \emph{the little co-group of $\bs{k}$}. The function 
	\begin{equation}
	\theta(R_i,R_j) = \exp{(-i\bs{g}_i\cdot\bs{t}_j)}
	\label{eq:fs}
	\end{equation}
is called the \emph{factor system} of the representation and it is completely fixed by the lattice symmetries. A derivation of equation (\ref{eqn:projrec}) can be found e.g~ in Ref.~\cite{Bradley:1972}. We offer a condensed review of the derivation in Appendix~\ref{app:groups1}. 

In many situations, the factor system Eq.~\eqref{eq:fs} is trivial. This is the case, especially,
\begin{itemize}
\item[($i$)] for symmorphic lattices because all $\bs{t}_j$ are zero,
\item[($ii$)] for momenta $\bs{k}$ \emph{inside} the Brillouin zone because all $\bs{g}_i$ are zero. 
\end{itemize}
In these situations, the definition (\ref{eqn:projrec}) reduces to that of an ``ordinary" representation. On the other hand, if the factor system is non-trivial, Eq.~\eqref{eqn:projrec} defines a \emph{projective representation}. This situation arises for $\bs{k}$-points on the surface of the Brillouin zone of non-symmorphic lattices. Interestingly, irreducible projective representations of a group can be higher-dimensional than their ordinary counterparts. 

This section is structured in the following way. In subsection~\ref{subsec:X227} we explain how the symmetries of the space group of the symmetric pyrochlore lattice (\# 227) protect a fourfold degeneracy at the $\mathrm{X}$-point of the Brillouin zone and why the spectrum disperses linearly around it, leading to a Dirac node. In subsection~\ref{subsec:X216} we show how breaking inversion symmetry splits the fourfold degeneracy into two linearly dispersing twofold degeneracies at different energies. The lower band of the upper representation crosses the upper band of the lower representation along a 2D sheet in momentum space. In subsection~\ref{subsec:WeylAppear} we explain why this crossing gaps out everywhere except of four points where it is protected by lattice symmetries. These four points are the Weyl nodes of the WSM phase in our model.


\subsection{The $\textrm{X}$ point in the symmetric pyrochlore lattice}\label{subsec:X227}


\subsubsection{Summary}

We will now analyze in more detail how the symmetries of the non-symmorphic pyrochlore lattice protect the Dirac nodes at the $\mathrm{X}$ point, where the non-trivial application of Eq.~\eqref{eqn:projrec} arises. Taking the spin-orbit coupling into account, the little co-group $\overline{G}^{\textrm{X}}$ contains 32 elements and is isomorphic to the double-valued $D_{4h}$. According to the tables in Ref.~\cite{Bradley:1972}, the \emph{only} irreducible projective representation compatible with the factor system is \emph{four-dimensional}. This already points to the four-fold degeneracy observed in Fig.~\ref{fig:spectra}(a). It also implies that at a commensurate filling with $2+4n$ ($n\in \mathbb{N}_0$) electrons per unit cell, a band insulator is {\em not} possible. This result complements a similar result found in Ref.~\cite{Parameswaran:2013} for the case of a vanishing spin-orbit coupling but with arbitrary electron-electron interaction.

By considering the generators of $\overline{G}^{\textrm{X}}$, it is possible to find the factor system and a symmetry-adapted basis that spans the 4DIR. This is achieved in Eq.~\eqref{eqn:sub:Xfunctions}. We further construct the effective low-energy Hamiltonian, given by Eq.~\eqref{eqn:227XHam-afterT}, which demonstrates that the spectrum indeed disperses linearly around the 4DIR.

\subsubsection{The factor system}
\begin{figure}[b!]
  \begin{center}
	\includegraphics[width=0.3\textwidth]{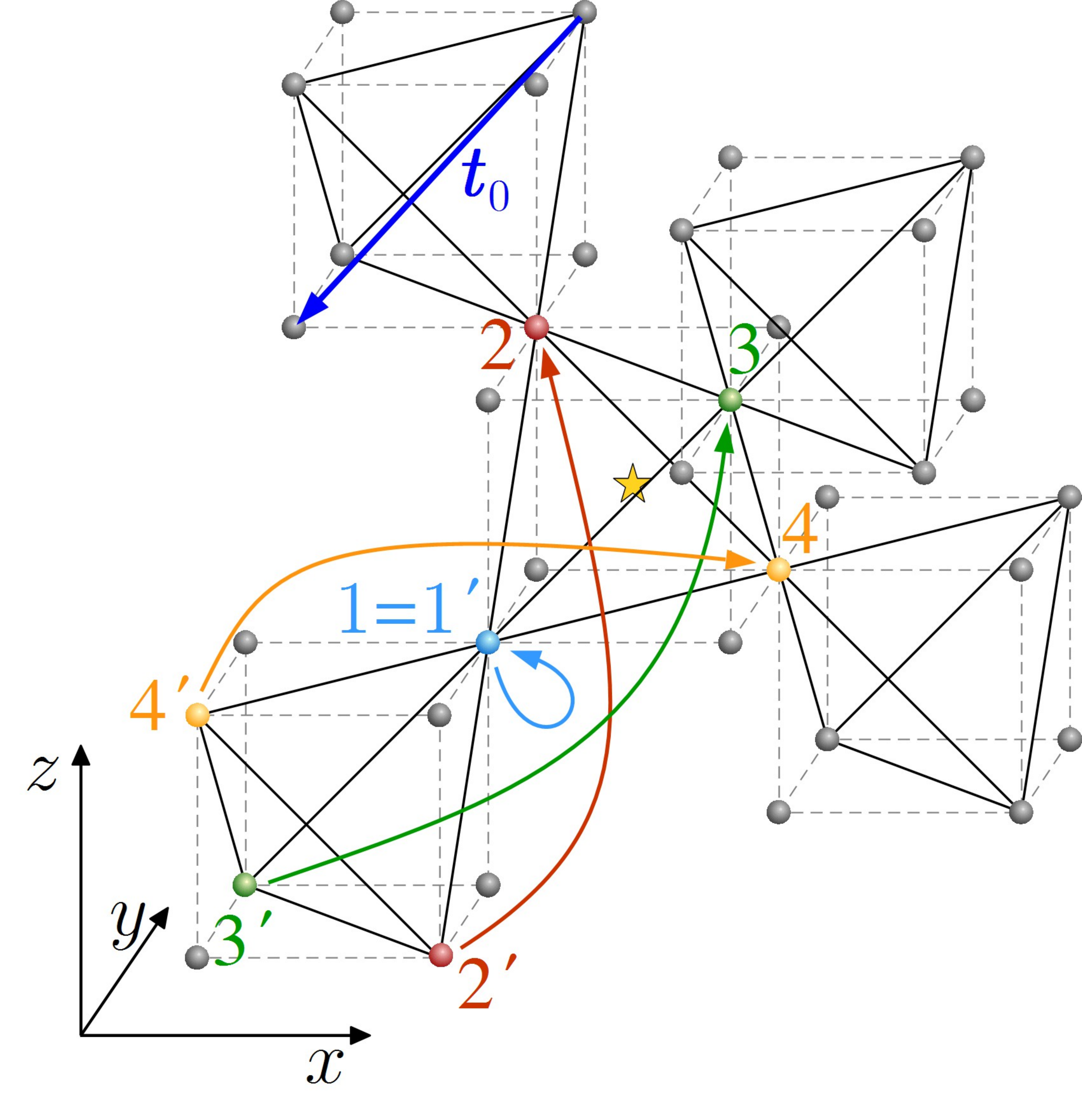}   		
  \end{center}
  \caption{(Color online) The action of $\cosrepnb{\mathcal{I}}{\bs{t}_0}$, where $\mathcal{I}$ is the inversion around the centre of symmetry indicated by the star, and $\bs{t}_0$ is a non-Bravais vector shift indicated by the blue arrow in the upper left tetrahedron. This symmetry operation maps the primed sites to the non-primed ones and is equivalent to the space inversion around site $1$. Note that sites $1$ and $2$ do not change their $x$-coordinate under the transformation while sites $3$ and $4$ are shifted by $-2a$. This means that $\cosrepnb{\mathcal{I}}{\bs{t}_0}$ preserves the amplitude of Bloch wave functions with $\bs{k}=\bs{X}=\frac{\pi}{2a}(1,0,0)$ on sites $1,2$, but it \emph{changes} the amplitude on sites $3,4$ by a factor $\exp{(\mathrm{i} \frac{\pi}{2a}\cdot 2a)= -1}$. The non-uniformity of these factors is a consequence of the non-symmorphicity of the symmetry operation.
  }
  \label{fig:minus-signs}
\end{figure}

To be specific, we consider in the following the $\textrm{X}$ point with coordinates $\bs{X}=\frac{\pi}{2a}(1,0,0)$. The little co-group $\overline{G}^{\textrm{X}}$ is generated by point operations $\mathcal{I}$, $C_{2z}^+$ and $C_{4x}^+$ where $C_{ni}^+$ is an $n$-fold rotation around axis $i$ in the positive (counter-clockwise) direction and the centre of symmetry is the centre of any (but fixed) tetrahedron. Of these generators,
\begin{itemize}
\item $\mathcal{I}$ and $C_{4x}^+$ are associated with a non-Bravais lattice shift $\boldsymbol{t}_0=-a(1,1,1)$,
\item $\mathcal{I}$ and $C_{2z}^+$ transform the $\textrm{X}$ point to an equivalent point displaced by $\boldsymbol{g}=-\frac{\pi}{a}(1,0,0)$.
\end{itemize}
With this information and knowing that $\exp{(-\mathrm{i}\boldsymbol{g}\cdot\boldsymbol{t}_0)}=-1$, it is easy to find the factor system between the group generators listed in Table~\ref{tab:Xfacsys}. 
	\begin{table}
	\begin{tabular}{rc|ccc}
	$\,$		&		$\,$	& 	$\mathcal{I}$&	$C_{2z}^+$		&	$C_{4x}^+$ 			\\
	$\,$		& 		$\,$	&	$(\bs{t}_0)$& $(\bs{0})$		& $(\bs{t}_0)$			\\ 	\hline
	$\mathcal{I}$&	$(\bs{g})$	&	$-1$		&	$+1$			&	$-1$				\\
	$C_{2z}^+$	&	$(\bs{g})$	&	$-1$		&	$+1$			&	$-1$				\\
	$C_{4x}^+$	&	$(\bs{0})$	&	$+1$		&	$+1$			&	$+1$				\\
	\end{tabular}
	\caption{Factor system of the generators of the little co-group $\overline{G}^\textrm{X}$. The vector indicated next to each point operation in the column is the non-Bravais lattice shift associated with it. The vector indicated next to it in the row gives the reciprocal lattice shift of the $\textrm{X}$ point under the point operation.}
	\label{tab:Xfacsys}
	\end{table}

An important consequence of the non-trivial factor system in Table~\ref{tab:Xfacsys} is that certain commuting point-symmetry operations are represented by \emph{anti}commuting operators and vice versa. We illustrate this fact with the relation between space inversion $\mathcal{I}$ and the two-fold rotation $C_{2z}^+$. These symmetry operations commute, i.e.
	\begin{subequations}\label{eqn:sign-ex}
	\begin{equation}
	\mathcal{I}\circ C_{2z}^+ = C_{2z}^+ \circ \mathcal{I}.\label{eqn:sign-ex1}
	\end{equation}
However, from Table~\ref{tab:Xfacsys} it follows that their (projective) representations {\em anticommute}:
	\begin{equation}
	\overline{D}_\textrm{X}(\mathcal{I})\overline{D}_\textrm{X}(C_{2z}^+) = -\overline{D}_\textrm{X}(C_{2z}^+)\overline{D}_\textrm{X}(\mathcal{I}).\label{eqn:sign-ex2}
	\end{equation}
	\end{subequations}
The origin of Eq.~\eqref{eqn:sign-ex2} can be traced back to the non-symmorphicity of the pyrochlore lattice. As shown in Fig.~\ref{fig:minus-signs}, inversion $\{\mathcal{I},\bs{t}_0\}$ maps three out of four sites to neighboring unit cells. As a result, when inversion acts on Bloch wave functions at X, non-uniform exponential factors $\exp{\left(\mathrm{i}\bs{X}\cdot\bs{r}\right)}$ have to be taken into account. Indeed, in the convention of Fig.~\ref{fig:minus-signs}, the amplitudes at sites $1,2$ are unchanged while the amplitudes at sites $3,4$ acquire a factor 
	\begin{equation}
	\exp{\left(\mathrm{i}\bs{X}\cdot 2\bs{b}_{13}\right)} = -1 = \exp{\left(\mathrm{i}\bs{X}\cdot 2\bs{b}_{14}\right)}.
	\end{equation}
The operation $C_{2z}^+$ exchanges sites $1,2$ with sites $3,4$ i.e. those that change sign under $\overline{D}_{\mathrm{X}}(\mathcal{I})$ with those that do not. This is just the statement of Eq.~\eqref{eqn:sign-ex2}.


\subsubsection{Construction of a symmetry adapted basis}

We can use our formalism to deduce the existence of the fourfold degeneracy at the $\mathrm{X}$ point by constructing a symmetry-adapted basis. Note that the representation $\overline{D}_{\textrm{X}}(C_{2x}^+)=\overline{D}_{\textrm{X}}(C_{4x}^+)^2$ \emph{commutes} with the representation $\overline{D}_{\textrm{X}}(\mathcal{I})$ and that they both commute with $\mathcal{H}(\textrm{X})$. Hence, a basis can be found that diagonalizes these three operators simultaneously. Because $\overline{D}_{\mathrm{X}}(\mathcal{I})^2=\overline{D}_{\textrm{X}}(C_{2x}^+)^2=-\mathbb{1}$ (see Table~\ref{tab:Xfacsys}), both $\overline{D}_{\textrm{X}}(\mathcal{I})$ and $\overline{D}_{\textrm{X}}(C_{2x}^+)$ have eigenvalues $\pm\mathrm{i}$. Let
	\begin{subequations}
	\label{eqn:sub:Xfunctions}
	\begin{equation}
	\ket{\psi_1}\label{eqn:Xfunctions1}
	\end{equation}
be an eigenvector of $\mathcal{H}(X)$ with eigenvalue $A=\pm \mathrm{i}$ under $\overline{D}_{\textrm{X}}(\mathcal{I})$ and eigenvalue $B=\pm\mathrm{i}$ under $\overline{D}_{\textrm{X}}(C_{2x}^+)$. Then, the states
	\begin{align}
	\ket{\psi_2} =& \overline{D}_{\textrm{X}}(C_{2z}^+)\ket{\psi_1} \\
	\ket{\psi_3} =& \overline{D}_{\textrm{X}}(C_{4x}^+)\ket{\psi_1} \\
	\ket{\psi_4} =& \overline{D}_{\textrm{X}}(C_{2z}^+) \overline{D}_{\textrm{X}}(C_{4x}^+)\ket{\psi_1} 
	\end{align}
	\end{subequations}
are eigenvectors at the same energy with {\em distinct} eigenvalues under $\overline{D}_{\textrm{X}}(\mathcal{I})$ and $\overline{D}_{\textrm{X}}(C_{2x}^+)$, as indicated in Table~\ref{tab:Xdeg}. Hence, the four states (\ref{eqn:sub:Xfunctions}) are mutually orthogonal and span the (projective) 4DIR at the $\mathrm{X}$ point. We also see that the 4DIR splits, if any of the symmetry elements $\cosrepnb{I}{\boldsymbol{t}_0}$, $\cosrepnb{C_{4x}^+}{\boldsymbol{t}_0}$ and $\cosrep{C_{2z}^+}{0}$ is removed from the space group. This agrees with findings of Ref.~\cite{Yang:2014b} that a rotation symmetry is an essential ingredient to obtain a protected Dirac node at a time-reversal invariant momentum. Note that the presence of time-reversal $\mathcal{T}$ is not relevant for the existence of the 4DIR.

	\begin{table}
	\begin{tabular}{r|c|c|c|c}
	$\,$					&	$\ket{\psi_1}$	&	$\ket{\psi_2}$	&	$\ket{\psi_3}$	&	$\ket{\psi_4}$	\\ \hline
	$\overline{D}_{\textrm{X}}(\mathcal{I})$				&	$+A$	&	$-A$	&	$-A$	&	$+A$	\\
	$\overline{D}_{\textrm{X}}(C_{2x}^+)$					&	$+B$	&	$-B$	&	$+B$	&	$-B$	\\
	\end{tabular}
	\caption{Eigenvalues of the states (\ref{eqn:sub:Xfunctions}) under $\overline{D}_{\textrm{X}}(\mathcal{I})$ and $\overline{D}_{\textrm{X}}(C_{2x}^+)$. $A$ and $B$ can both assume values  $\pm\mathrm{i}$.}
	\label{tab:Xdeg}
	\end{table}
%


\subsubsection{Linear dispersion}

We further derive the linear dispersion of the spectrum around the $\textrm{X}$ point. A formal group theoretical argument, also given in Ref.~\cite{Young:2012}, is expounded in Appendix~\ref{sec:sym-prod}. Here, we instead present a reasoning that explicitly shows the role of symmetries. 

Insight can be gained by expanding the Hamiltonian perturbatively in the momentum $ \boldsymbol{p}$ around the X-point. The first order term is
	\begin{eqnarray}
	\left[\mathcal{H}^{\textrm{X}}_{\textrm{pert.}}(\bs{p})\right]_{ij} &=& \bra{\psi_i}\left(\left.\pder{\mathcal{H}(\boldsymbol{k})}{\boldsymbol{k}}\right|_\textrm{X}\cdot \boldsymbol{p}\right)\ket{\psi_j}\nonumber\\
	&\approx&\bra{\psi_i}\left[\mathcal{H}(\textrm{X}+\bs{p})-\mathcal{H}(\textrm{X})\right]\ket{\psi_j}.\label{eqn:X-PT}
	\end{eqnarray}
If there is a direction of ${\bs p}$ for which some of the matrix elements are non-zero, the spectrum might disperse linearly in that direction. Otherwise, the spectrum disperses at least quadratically in any direction.

Symmetries pose constraints on the matrix elements of Eq.~\eqref{eqn:X-PT}. To find them, we use the transformation laws of the Bloch functions (\ref{eqn:sub:Xfunctions}) deduced from Tables~\ref{tab:Xfacsys} and~\ref{tab:Xdeg} together with the fact that the perturbation Hamiltonian transforms according to a vector representation,
	\begin{equation}
	R: \left(\left.\pder{\mathcal{H}(\boldsymbol{k})}{\boldsymbol{k}}\right|_\textrm{X}\cdot \boldsymbol{p} \right) \mapsto \left( \left.\pder{\mathcal{H}(\boldsymbol{k})}{\boldsymbol{k}}\right|_\textrm{X}\cdot \left(R \boldsymbol{p}\right) \right),\label{eq:HR}	
	\end{equation}
under a point operation $R \in \overline{G}^\textrm{X}$. Equation~\eqref{eq:HR} follows from $R: \mathcal{H}(\textrm{X}+\bs{p})\mapsto \mathcal{H}(R\textrm{X}+R\bs{p})=\mathcal{H}(\textrm{X}+R\bs{p})$.

We start by considering $\mathcal{I}$, which flips the sign of all components of the vector $\boldsymbol{p}$. If the Bloch functions $\bra{\psi_i}$ and $\ket{\psi_j}$ have opposite eigenvalues of $\overline{D}_{\mathrm{X}}(\mathcal{I})$, $+A$ and $-A$, the corresponding matrix element (\ref{eqn:X-PT}) is transformed to minus itself under $\mathcal{I}$ and hence must be zero. This reasoning forces the matrix elements indicated in Table~\ref{tab:X-PT} by the crossed font ``$\cancel{p_i}$" to vanish.

We further consider the operation $C_{2x}^+$ which flips the sign of $p_y$ and $p_z$ and preserves the sign of $p_x$. If the Bloch functions $\bra{\psi_i}$ and $\ket{\psi_j}$ have \emph{the same} eigenvalue $\pm B$ under $C_{2x}^+$, they produce a factor $(\pm B)^2=-1$ under that transformation. This means that the corresponding $p_x$ matrix element maps to minus itself and must vanish. On the other hand, if the two Bloch functions have \emph{opposite} eigenvalues, they produce a factor of $+1$ and the corresponding $p_y$ and $p_z$ matrix elements are forced to be zero. Both of these constraints are indicated by the back-crossed terms ``$\bcancel{p_i}$" in Table~\ref{tab:X-PT}.
	\begin{table}
	\begin{tabular}{r|c|c|c|c}
	$\,$&	$\ket{\psi_1}$ 	
		&	$\ket{\psi_2}$	
		&	$\ket{\psi_3}$	
		&	$\ket{\psi_4}$	\\ 
	$\,$&	$(+A,+B)$ 	
		&	$(-A,-B)$	
		&	$(-A,+B)$	
		&	$(+A,-B)$ \\ \hline
	$\bra{\psi_1}\:(-A,-B)$	
		&	${\color{red}\cancel{p_x}},\color{OliveGreen}\xcancel{p_y},\color{OliveGreen}\xcancel{p_z}$	
		&	${\color{blue}\bcancel{p_x}},{\color{black}{p_y}},{\color{black}{p_z}}$	
		&	${\color{black}{p_x}},{\color{blue}\bcancel{p_y}},{\color{blue}\bcancel{p_z}}$	
		&	${\color{OliveGreen}\xcancel{p_x}},{\color{red}\cancel{p_y}},{\color{red}\cancel{p_z}}$	\\ \hline
	$\bra{\psi_2}\:(+A,+B)$	
		&	${\color{blue}\bcancel{p_x}},{\color{black}{p_y}},{\color{black}{p_z}}$	
		&	${\color{red}\cancel{p_x}},{\color{OliveGreen}\xcancel{p_y}},{\color{OliveGreen}\xcancel{p_z}}$	
		&	${\color{OliveGreen}\cancel{p_x}},{\color{red}\xcancel{p_y}},{\color{red}\cancel{p_z}}$	
		&	${\color{black}{p_x}},{\color{blue}\bcancel{p_y}},{\color{blue}\bcancel{p_z}}$	\\ \hline
	$\bra{\psi_3}\:(+A,-B)$	
		&	${\color{black}{p_x}},{\color{blue}\bcancel{p_y}},{\color{blue}\bcancel{p_z}}$	
		&	${\color{OliveGreen}\xcancel{p_x}},{\color{red}\cancel{p_y}},{\color{red}\cancel{p_z}}$	
		&	${\color{red}\cancel{p_x}},{\color{OliveGreen}\xcancel{p_y}},{\color{OliveGreen}\xcancel{p_z}}$	
		&	${\color{blue}\bcancel{p_x}},{\color{black}{p_y}},{\color{black}{p_z}}$	\\ \hline
	$\bra{\psi_4}\:(-A,+B)$	
		&	${\color{OliveGreen}\xcancel{p_x}},{\color{red}\cancel{p_y}},{\color{red}\cancel{p_z}}$	
		&	${\color{black}{p_x}},{\color{blue}\bcancel{p_y}},{\color{blue}\bcancel{p_z}}$	
		&	${\color{blue}\bcancel{p_x}},{\color{black}{p_y}},{\color{black}{p_z}}$	
		&	${\color{red}\cancel{p_x}},{\color{OliveGreen}\xcancel{p_y}},{\color{OliveGreen}\xcancel{p_z}}$
	\end{tabular}
	\caption{The two terms in brackets indicate the eigenvalues of the states under $\overline{D}_{\textrm{X}}(\mathcal{I})$ and $\overline{D}_{\textrm{X}}(C_{2x}^+)$, respectively. The crossed ``$\cancel{p_i}$" terms [shown in blue and green in the online version] indicate matrix elements (\ref{eqn:X-PT}) that vanish due to $\mathcal{I}$ and the back-crossed terms ``$\bcancel{p_i}$" [shown in blue and green in the online version] indicate those that vanish due $C_{2x}^+$ symmetry. 
	Since all $p_x$,$p_y$ and $p_z$ remain uncrossed for some pair of states, the spectrum disperses linearly in all directions around the $\textrm{X}$ point.}
	\label{tab:X-PT}
	\end{table}

The remaining generators of the little co-group do not force any of the remaining matrix elements to be zero. Since there are $p_x$, $p_y$ and $p_z$ terms uncrossed for some pair of wave functions in Table~\ref{tab:X-PT}, the spectrum disperses linearly in all directions. 


\subsubsection{Effective Dirac Hamiltonian}

To deduce the form of the effective Hamiltonian, we also analyze how symmetries relate the non-vanishing matrix elements in Eq.~\eqref{eqn:X-PT}. First, the rotation $C_{2z}^+$ leads to
	\begin{subequations}\label{eqn:sub:227Xpart1}
	\begin{eqnarray}
	\left[\mathcal{H}^{\textrm{X}}_{\textrm{pert.}}(p_x)\right]_{13} &=& -\left[\mathcal{H}^{\textrm{X}}_{\textrm{pert.}}(p_x)\right]_{24} \\ 
	\left[\mathcal{H}^{\textrm{X}}_{\textrm{pert.}}(p_y)\right]_{12} &=& +\left[\mathcal{H}^{\textrm{X}}_{\textrm{pert.}}(p_y)\right]_{21} \\ 
	\left[\mathcal{H}^{\textrm{X}}_{\textrm{pert.}}(p_z)\right]_{12} &=& -\left[\mathcal{H}^{\textrm{X}}_{\textrm{pert.}}(p_z)\right]_{21} 
	\end{eqnarray}
	\end{subequations}
and the same relations with $(1,2,3,4)\leftrightarrow(3,4,1,2)$. Second, the remaining little co-group generator $C_{4x}^+$, which maps $(p_x,p_y,p_z)$ to $(p_x,-p_z,p_y)$, leads to
	\begin{subequations}\label{eqn:sub:227Xpart2}
	\begin{eqnarray}
	\left[\mathcal{H}^{\textrm{X}}_{\textrm{pert.}}(p_y)\right]_{12} &=& (-B)\left[\mathcal{H}^{\textrm{X}}_{\textrm{pert.}}(p_z)\right]_{34} \\ 
	\left[\mathcal{H}^{\textrm{X}}_{\textrm{pert.}}(p_z)\right]_{12} &=& (+B)\left[\mathcal{H}^{\textrm{X}}_{\textrm{pert.}}(p_y)\right]_{34} \\ 
	\left[\mathcal{H}^{\textrm{X}}_{\textrm{pert.}}(p_x)\right]_{13} &=& (+B)\left[\mathcal{H}^{\textrm{X}}_{\textrm{pert.}}(p_x)\right]_{31} \
	\end{eqnarray}
	\end{subequations}
and \emph{opposite sign} relations with $(1,2,3,4)\leftrightarrow(4,3,2,1)$. Combining these results with hermiticity and time-reversal leads to an effective Dirac Hamiltonian 
	\begin{equation}
	\mathcal{H}^{\textrm{X}}_{\textrm{Dirac}}(\bs{p}) = \varepsilon(\textrm{X})\mathbb{1} + a p_x \Gamma_1 + b\left(p_y \Gamma_2 + p_z \Gamma_3 \right)\label{eqn:227XHam-afterT}
	\end{equation}
with $a,b\in\mathbb{R}$. The Dirac matrices $\Gamma_i$ in our basis are given by 
	\begin{subequations}\label{eqn:KdotPdirac}
	\begin{eqnarray}
		\Gamma_1 &=& \frac{1}{\sqrt{2}} \left(\sigma_x + \mathrm{i} B\sigma_y\right)\otimes \sigma_z \\
		\Gamma_2 &=& \mathbb{1}\otimes \sigma_x \\
		\Gamma_3 &=& \mathrm{i}B \,\mathbb{1}\otimes\sigma_y \\
		\Gamma_4 &=& \sigma_z \otimes \sigma_z
	\end{eqnarray}
	\end{subequations}
and fulfill the anticommutation relation
	\begin{equation}
	\left\{ \Gamma_i,\Gamma_j \right\} = 2\delta_{ij}.\label{eqn:DiracHam}
	\end{equation}
The first set of the Pauli matrices in Eq.~\eqref{eqn:KdotPdirac} operates on the $2\times 2$ blocks of the matrix Hamiltonian (\ref{eqn:227XHam-afterT}), and the second set acts within these blocks. Diagonalizing the Dirac Hamiltonian \eqref{eqn:227XHam-afterT} leads to a spectrum with two doubly degenerate linearly dispersing bands
	\begin{equation}
	\varepsilon(\textrm{X} + \bs{p}) = \varepsilon(\textrm{X}) \pm \sqrt{(a p_x)^2 + b^2 (p_y^2 + p_z^2)}.
	\end{equation}
%


\subsection{The $\textrm{X}$ point in the breathing pyrochlore lattice}\label{subsec:X216}

Breaking of the inversion symmetry decreases the allowed degeneracy at the $\textrm{X}$ point from $4$ to $2$, i.e., the 4DIR splits into two 2DIRs with the chemical potential in-between. The dispersion around each of these nodes is linear within the square face of the BZ and quadratic in the perpendicular direction. As discussed in Sec.~\ref{subsec:WeylAppear}, the Weyl nodes appear as symmetry protected crossings of the up-dispersing band of the lower 2DIR with the down-dispersing band of the upper 2DIR.

More specifically, the little co-group at the $\textrm{X}$ point of the $\mathcal{I}$-broken pyrochlore lattice is isomorphic to the double-valued $D_{2d}$ crystallographic point group. It is generated by the rotation $C_{2z}^+$ and the improper rotation
	\begin{equation}
	S_{4x}^-=\mathcal{I}\circ C_{4x}^+:(x,y,z)\mapsto (-x,z,-y ),
	\end{equation}
that satisfies $\left(S_{4x}^-\right)^2=C_{2x}^+$. Note that the $\mathcal{I}$-broken pyrochlore lattice is symmorphic, so the factor system is trivial and the degeneracy is given by the ``ordinary" IRs. According to~\cite{Bradley:1972}, all IRs that produce a minus sign under a $2\pi$-rotation are two-dimensional.

Let us now construct the states that span the 2DIR. Note that it is possible to simultaneously diagonalize the Hamiltonian $\mathcal{H}(\textrm{X})$, and operators $\overline{D}_{\textrm{X}}(S_{4x}^-)$ and $\overline{D}_{\textrm{X}}(C_{2x}^+)$. Let
	\begin{subequations}
	\label{eqn:sub:X216states}
	\begin{equation}
	\ket{\phi_1^\pm}
	\end{equation}
be an eigenstate of $\mathcal{H}(\textrm{X})$ with energy $\varepsilon(\textrm{X})\pm\Delta/2$ and an eigenstate of $\overline{D}_{\textrm{X}}(C_{2x}^+)$ with eigenvalue $B=\pm\mathrm{i}$. It follows that
	\begin{equation}
	\ket{\phi_2^\pm}=\overline{D}_\textrm{X}(C_{2z}^+)\ket{\phi_1}
	\end{equation}
	\end{subequations}
has the same energy as $\ket{\phi_1}$ but the opposite eigenvalue of $\overline{D}_\textrm{X}(C_{2x}^+)$. This implies that the two states are orthogonal and span the 2DIR at the $\mathrm{X}$ point. If an appropriate phase of $\ket{\psi_1}$ in (\ref{eqn:Xfunctions1}) is adopted, then
	\begin{equation}
	\ket{\phi_{1,2}^\pm} = \frac{1}{\sqrt{2}}\left(\ket{\psi_{1,2}} \pm \mathrm{i}\sqrt{B}\ket{\psi_{3,4}}\right) 
	\end{equation}
where $+(-)$ sign refers to the upper (lower) 2DIR, assuming that $\Delta > 0$, and the branch cut of the square root is along the negative real axis. As a consequence, in linear order in $\Delta$ and $\bs{p}$, the Dirac Hamiltonian~\eqref{eqn:227XHam-afterT} acquires an additional term due to the inversion-symmetry breaking,
\begin{equation}
\mathcal{H}^{\textrm{X}}_{\textrm{Dirac}}(\bs{p})\mapsto \mathcal{H}^{\textrm{X}}_{\textrm{Dirac}}(\bs{p})+\frac{\Delta}{2}\Gamma_{14},
\label{eq:Dirac-split}
\end{equation}
where $\Gamma_{ij}=-\frac{\mathrm{i}}{2}\left[\Gamma_i,\Gamma_j\right]$. The effect of such a term on the Dirac Hamiltonian has been discussed in Refs.~\cite{Burkov:2011b,Halasz:2012} and agrees with our observations, e.g.~in Fig.~\ref{fig:spectra}(b) and Eq.~\eqref{eqn:DiracHam2} below.

In analogy with the previous subsection, we can also investigate the dispersion of the 2DIRs around the $\mathrm{X}$ point using the symmetries. They again lead to certain constraints on the matrix elements between $\ket{\phi_1}$ and $\ket{\phi_2}$: If $\bra{\phi_i}$ and $\ket{\phi_j}$ have the \emph{same} eigenvalue $\pm B$ under $\overline{D}_{\mathrm{X}}(C_{2z}^x)$, the corresponding $p_x$ matrix element must vanish. If their eigenvalues are \emph{opposite}, the $p_y$ and $p_z$ elements are forced to be zero. These matrix elements are indicated in Table~\ref{tab:X-PT-2} by the crossed ``$\cancel{p_i}$" terms. On the other hand, the improper rotation $S_{4x}^-$ flips the sign of $p_x$, hence the matrix elements $\bra{\phi_i}p_x\ket{\phi_i}$ are mapped to minus themselves under the transformation and have to vanish. They are indicated in Table~\ref{tab:X-PT-2} by the back-crossed ``$\bcancel{p_i}$" terms. 
	\begin{table}
	\begin{tabular}{rc|c|c}
	$\,$& $\,$	&	$\ket{\phi_1}$ 	
				&	$\ket{\phi_2}$	\\
	$\,$&	$\,$&	$(+B)$ 	
				&	$(-B)$ 		\\ \hline
	$\bra{\phi_1}$ 	&$(-B)$	
		&	${\color{RedOrange}\bcancel{p_x}},\color{blue}\cancel{p_y},\color{blue}\cancel{p_z}$	
		&	${\color{blue}\cancel{p_x}},{\color{black}{p_y}},{\color{black}{p_z}}$	\\ \hline
	$\bra{\phi_2}$ 	&$(+B)$	
		&	${\color{blue}\cancel{p_x}},{{p_y}},{{p_z}}$	
		&	${\color{RedOrange}\bcancel{p_x}},{\color{blue}\cancel{p_y}},{\color{blue}\cancel{p_z}}$
	\end{tabular}
	\caption{The $\pm B$ terms indicate the eigenvalue of the state under a $C_{2x}^+$ rotation, which is the square of the eigenvalue under the improper $S_{4x}^-$ rotation. The blue crossed ``$\color{blue}\cancel{p_i}$" terms indicate matrix elements (\ref{eqn:X-PT}) that vanish due to $C_{2x}^+$ and the orange back-crossed ``$\color{RedOrange}\bcancel{p_i}$" terms indicate those that vanish due to $S_{4x}^-$. The uncrossed black terms are allowed by symmetry. The absence of $p_x$ implies that the spectrum disperses quadratically in this direction.}
	\label{tab:X-PT-2}
	\end{table}

The remaining matrix elements are non-zero. The improper rotation $S_{4x}^-$ relates
	\begin{subequations}
	\begin{eqnarray}
	\left[\mathcal{H}^{\textrm{X},\pm}_{\textrm{pert.}}(p_y)\right]_{12} &=& (-B) \left[\mathcal{H}^{\textrm{X},\pm}_{\textrm{pert.}}(p_z)\right]_{12} \\ 
	\left[\mathcal{H}^{\textrm{X},\pm}_{\textrm{pert.}}(p_y)\right]_{21} &=& (+B) \left[\mathcal{H}^{\textrm{X},\pm}_{\textrm{pert.}}(p_z)\right]_{21}
	\end{eqnarray}
	\end{subequations}
and the rotation $C_{2z}^+$ further leads to
	\begin{subequations}
	\begin{eqnarray}
	\left[\mathcal{H}^{\textrm{X},\pm}_{\textrm{pert.}}(p_y)\right]_{12} &=&  + \left[\mathcal{H}^{\textrm{X},\pm}_{\textrm{pert.}}(p_y)\right]_{21} \\ 
	\left[\mathcal{H}^{\textrm{X},\pm}_{\textrm{pert.}}(p_z)\right]_{12} &=& - \left[\mathcal{H}^{\textrm{X},\pm}_{\textrm{pert.}}(p_z)\right]_{21}.
	\end{eqnarray}
	\end{subequations}
Time reversal does not lead to further constraints. After considering the hermiticity, the perturbation Hamiltonian can be written as
	\begin{equation}
	\mathcal{H}^{\textrm{X},\pm}_{\textrm{eff.}}(\bs{p}) = \left(\varepsilon{(\textrm{X})}\pm \tfrac{\Delta}{2}\right)\mathbb{1} + b \left(p_y \sigma_x + \mathrm{i}B p_z \sigma_y \right) \label{eqn:DiracHam2}.
	\end{equation}
with $c\in\mathbb{R}$. The two bands disperse linearly along the $p_y$ and $p_z$ directions and  quadratically along the $p_x$ direction. This is in agreement with analysis of Eq. (\ref{eq:Dirac-split}) in Ref.~\cite{Burkov:2011b}.

 
\subsection{Appearance of the Weyl nodes}\label{subsec:WeylAppear}

The breaking of the inversion symmetry splits the linearly dispersing 4DIR discussed in subsection \ref{subsec:X227} into the two 2DIRs discussed in \ref{subsec:X216} that disperse linearly only within the square face of the Brillouin zone and quadratically along the $\Gamma\textrm{X}$ direction. The up-dispersing band of the lower 2DIR crosses the down-dispersing band of the upper 2DIR on a 2D sheet that locally looks like a one-sheet hyperboloid with the $\textrm{X}$ point in the centre. At a general ${\bs k}$-point, this crossing gaps out because there is no symmetry to protect it. Potential exceptions are ${\bs k}$-points with higher symmetry and we therefore discuss the $\textrm{XU}$ and the $\textrm{XW}$ lines in more details.

\subsubsection{The \textrm{XU}-line} 

Upon breaking the inversion symmetry, the four bands originating from the 4DIR at the $\textrm{X}$ point disperse as indicated in Fig.~\ref{fig:representations}(a). The question is whether the band crossing is protected or gapped. 

The little co-group $\overline{G}^{\textrm{XU}}$ is cyclic, the only generator is a mirror symmetry $\sigma_{d1}$. The possible eigenvalues of $\overline{D}_{\textrm{XU}}(\sigma_{d1})$ are $\pm\mathrm{i}$ and correspond to two different 1DIRs. Two bands with the \emph{same} eigenvalue can hybridize and their crossings generally gap out. On the other hand, two bands with \emph{opposite} eigenvalues cannot hybridize and hence their crossings are protected. To ascribe the appropriate 1DIR to each of the four bands illustrated in Fig.~\ref{fig:representations}(a), we devise the following two arguments.

First, if we restore the inversion symmetry, then $\mathcal{T}\circ \mathcal{I}$ enters the little co-group, and the bands 1 and 2 in Fig.~\ref{fig:representations}(a) become degenerate. This element commutes with $\sigma_{d1}$. Furthermore, both $\mathcal{T}\circ \mathcal{I}$ and $\sigma_{d1}$ are associated with zero shift $\bs{g}=\bs{0}$ in momentum space, leading to a trivial factor system, therefore
\begin{equation}
\left[\overline{D}_\textrm{XU}(\mathcal{T}\circ\mathcal{I}),\overline{D}_\textrm{XU}(\sigma_{d1}) \right] = 0.
\end{equation}
Note that $\overline{D}_\textrm{XU}(\mathcal{T}\circ\mathcal{I})$ is antiunitary because of the time-reversal. Now, if state $\ket{\psi_1}$ from band $1$ is eigenvector of $\overline{D}_{\textrm{XU}}(\sigma_{d1})$ with eigenvalue $B=\pm\mathrm{i}$, then the state $\ket{\psi_2}=\overline{D}_\textrm{XU}(\mathcal{T}\circ\mathcal{I})\ket{\psi_1}$ from band 2 has eigenvalue $B^*=-B$ under $\overline{D}_{\textrm{XU}}(\sigma_{d1})$ and hence belongs to the other representation. Similar conclusion can be found for bands 3 and 4.

Second, if we keep inversion symmetry broken and move along the $\textrm{XU}$ line to the $\textrm{X}$ point, a perpendicular mirror symmetry $\sigma_{d2}$ enters the little co-group that fulfils $\sigma_{d1}\circ \sigma_{d2} = \overline{E} \circ \sigma_{d2} \circ \sigma_{d1}$, where $\overline{E}$ is a $2\pi$-rotation. Therefore
\begin{equation}
\left\{\overline{D}_\textrm{X}(\sigma_{d1}),\overline{D}_\textrm{X}(\sigma_{d2}) \right\} = 0.
\end{equation}
The state $\ket{\psi_3}=\overline{D}_{\textrm{X}}(\sigma_{d2})\ket{\psi_1}$ from band $3$ has the same energy as $\ket{\psi_1}$ but has eigenvalue $-B$ under $\overline{D}_{\textrm{X}}(\sigma_{d1})$. This means that bands $1$ and $3$ belong to different 1DIRs. The same conclusion can be found for bands $2$ and $4$.

Our conclusions are summarized in Fig.~\ref{fig:representations}(a). Bands with eigenvalue $+B$ are indicated by dashed red lines, while those with eigenvalue $-B$ by solid blue lines. We see that the two crossing bands belong to the same representation, so the crossing will in general be gapped out by hybridization.

\begin{figure}[t!]
  \begin{center}
	\includegraphics[width=0.4\textwidth]{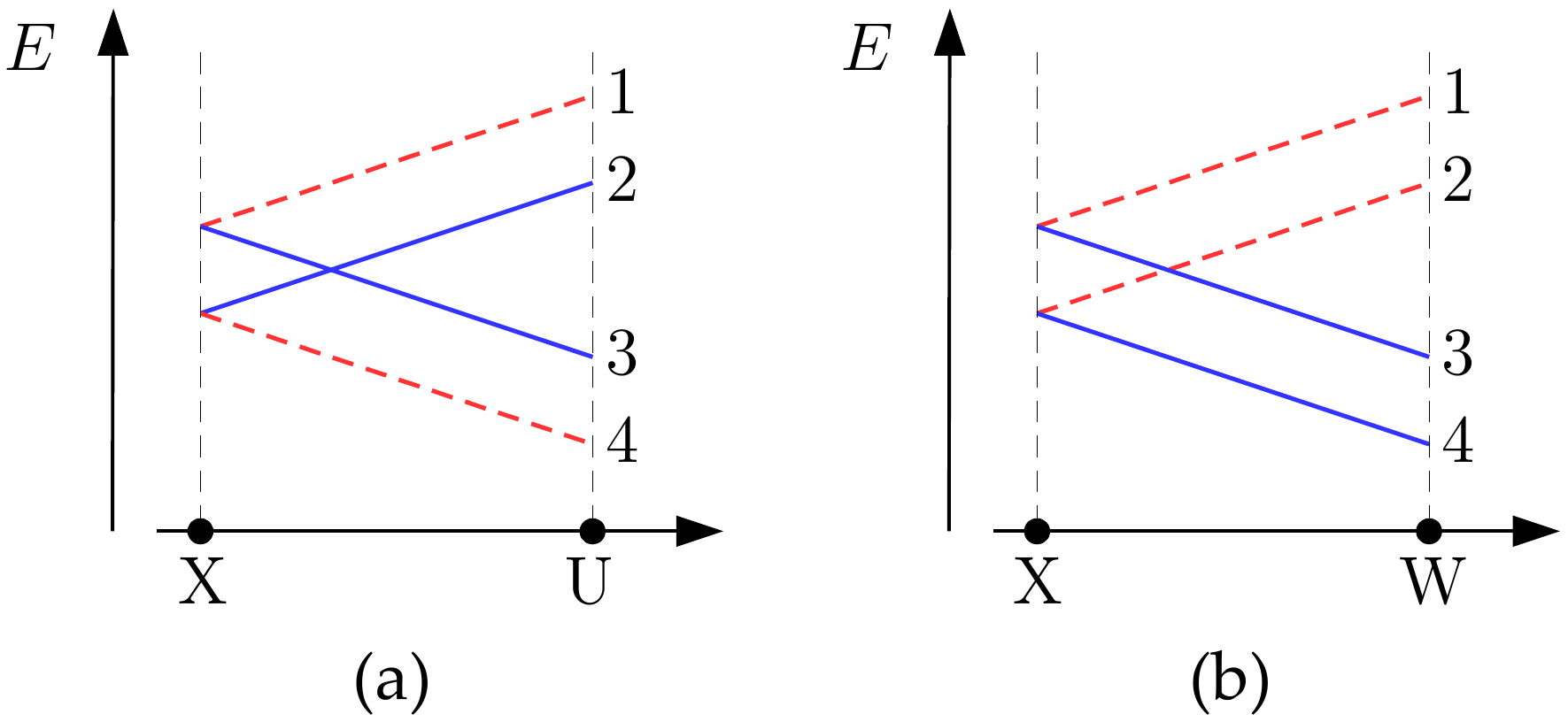}   		
  \end{center}
  \caption{(Color online) (a) A schematic sketch of the four bands along the $\textrm{XU}$ lines, originating from the 4DIR at the $\textrm{X}$ points. There are only two 1DIRs along this line. Bands belonging to one of them are indicated by dashed red lines, those belonging to the other representation are shown in solid blue. The bands that cross belong to the \emph{same} representation, hence the crossing gaps out. (b) Analogous analysis along the $\textrm{XW}$ line reveals that the crossing bands belong to \emph{different} representations. Such a crossing is protected by symmetry and yields a Weyl node that can be gapped out only by annihilation with another Weyl node.}
  \label{fig:representations}
\end{figure}

\subsubsection{The \textrm{XW}-line} 

We finally investigate the crossing along the $\textrm{XW}$ line. For concreteness, we consider the line parallel with the $z$-axis. The little co-group $\overline{G}^{\textrm{XW}}$ is generated by a two-fold rotation $C_{2z}^+$. Possible eigenvalues of $\overline{D}_{\textrm{XW}}(C_{2z}^+)$ are $\pm\mathrm{i}$ and correspond to two different 1DIRs. We want to assign a representation to each of the bands sketched in Fig.~\ref{fig:representations}(b). The argumentation proceeds again in two steps.

First, if we restore the inversion symmetry $\mathcal{I}$, then $\mathcal{T}\circ\mathcal{I}$ and mirror symmetries $\sigma_x$ and $\sigma_y$ appear in the little co-group, and bands $1$ and $2$ become degenerate. Furthermore, the system is again non-symmorphic and the factor system is non-trivial. A careful calculation reveals that representations $\overline{D}_{\textrm{XW}}(C_{2z}^+)$, $\overline{D}_{\textrm{XW}}(\sigma_x)$ and $\overline{D}_{\textrm{XW}}(\sigma_y)$ mutually commute, so a basis can be found that diagonalizes all of them simultaneously. 

Let $\ket{\psi_1}$ be an element of such a basis with eigenvectors $B=\pm\mathrm{i}, A_x=\pm 1, A_y=\pm\mathrm{i}$, respectively. Since 
	\begin{equation}
	\overline{D}_{\textrm{XW}}(\sigma_x) \overline{D}_{\textrm{XW}}(\sigma_y) = - \overline{D}_{\textrm{XW}}(C_{2z}^+)
	\end{equation}
the eigenvalues are constrained by $A_x A_y = -B$. It turns out that the state $\ket{\psi_2}=\overline{D}_{\textrm{XW}}(\mathcal{T}\circ\mathcal{I})\ket{\psi_1}$ from band $2$ has the same energy and eigenvalues $B, -A_x, -A_y$, respectively. The unchanged sign of the eigenvalue of $\overline{D}_{\textrm{XW}}(C_{2z}^+)$ means that bands 1 and 2 belong to the same representation. Analogous statement can be made about bands $3$ and $4$.

Second, if we move towards the $\textrm{X}$ point, then $C_{2y}^+$ enters the little co-group and bands $1$ and $3$ become degenerate. Since $C_{2x}^+ \circ C_{2y}^+ = \overline{E} \circ C_{2y}^+ \circ C_{2x}^+$, we find that 
\begin{equation}
\left\{\overline{D}_\textrm{X}(C_{2x}^+),\overline{D}_\textrm{X}(C_{2y}^+) \right\} = 0.
\end{equation}
This means that the state $\ket{\psi_3} = \overline{D}_{\textrm{X}}(C_{2y}^+)$ from band 3 is an eigenstate at $\textrm{X}$ with the same energy and eigenvalue $-B$ under $\overline{D}_{\textrm{X}}(C_{2z}^+)$ This implies that bands $1$ and $3$ in Fig.~\ref{fig:representations}(b) belong to different representations. Analogous conclusion can be derived for bands $2$ and $4$.

Note that the bands crossing along the $\textrm{XW}$ line belong to different representations. This means that the crossing is protected. It corresponds to one of the Weyl nodes of the Weyl semimetal phase in our model. The other Weyl points are related by crystal symmetries.


\section{$\{111\}$ and $\{11\bar{1}\}$ Surface states}\label{sec:surface}

\subsection{Terminations of a $(111)$ slab}
The topological nature of the WSM phase is manifest in the exotic surface states that have the form of \emph{Fermi arcs} and that are robust against all local perturbations~\cite{Wan:2011,Hosur:2013}. The endpoints of the Fermi arcs are given by projections of the bulk Weyl nodes onto the surface Brillouin zone (SBZ). In the following, we focus on the $\{111\}$ and $\{11\bar{1}\}$ surfaces as these are the natural cleavage planes of the pyrochlore oxides. Besides the topological surface states exhibited by the WSM of model (\ref{eqn:Hsimp}), we also identify non-topological surface states in the DSM and the INS phase for certain terminations of the pyrochlore lattice.

Along the $\{111\}$ directions, the pyrochlore lattice can be viewed as a stack of alternating layers of triangular and kagome lattices. Hence, the two simplest open boundary conditions correspond to terminating the crystal at sites making up either the kagome lattice (K) or the triangular lattice (T). Note that the kagome termination cuts the outermost tetrahedra. Because of the inequivalent bonds in the $\mathcal{I}$-broken state, we further have to specify whether the bonds connecting the outermost layer of sites to the next layer are strong (S) or weak (W). This gives four possible terminations illustrated in Fig.~\ref{fig:surf-terminations}. For the DSM phase the labels S and W are redundant and can be dropped. A suitable geometry for experimental studies is a thin slab with large lateral dimensions~\cite{Potter:2014}. If we set the $(111)$ surface on the top and we limit our attention to $\delta>0$, then only KS and TW can be realized on the top, and only KW and TS on the bottom surface, as shown in Fig.~\ref{fig:surf-terminations}.
\begin{figure}
  \begin{center}
	\includegraphics[width=0.475\textwidth]{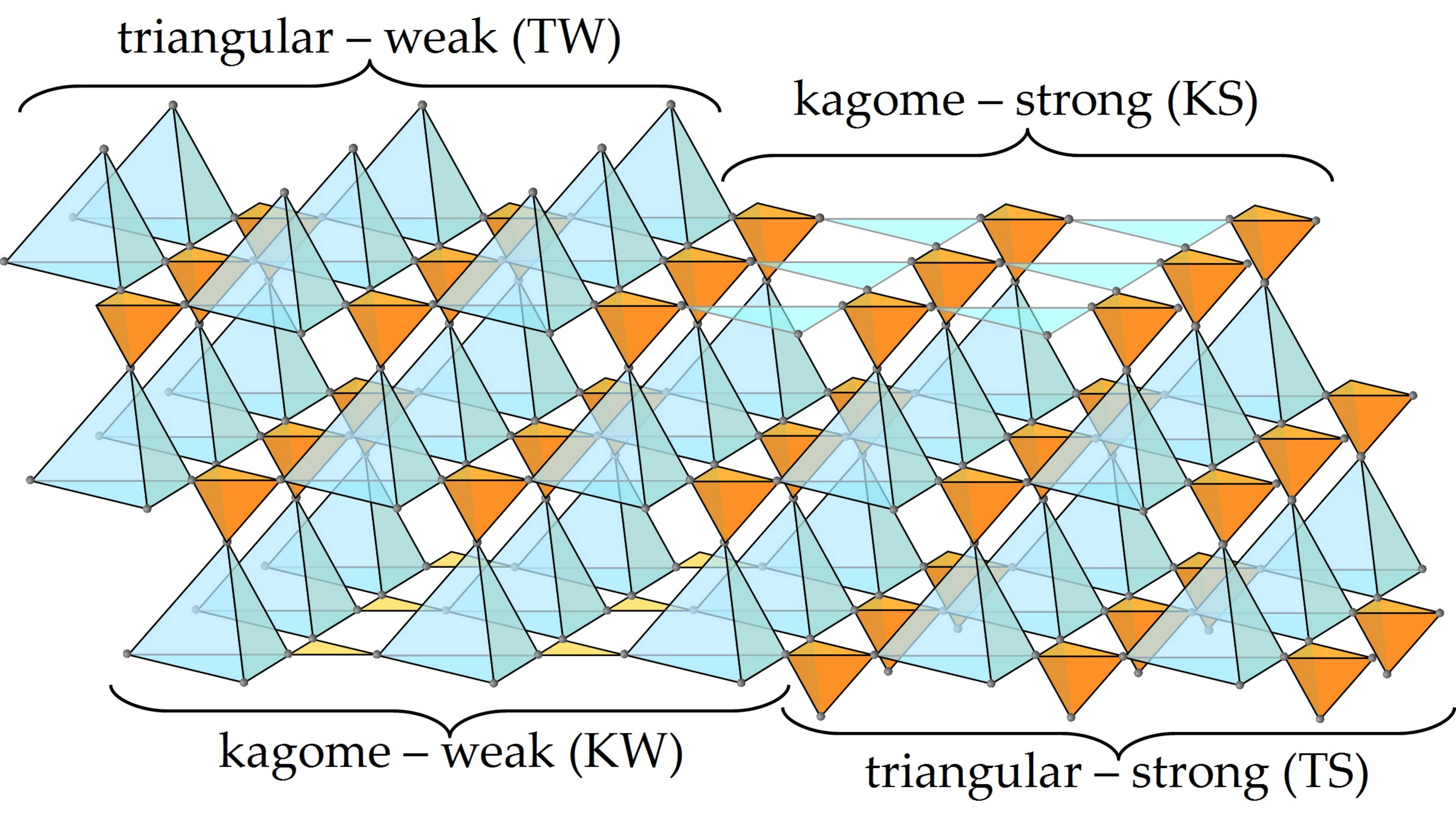}  		
  \end{center}
  \caption{(Color online) Illustration of the four crystal termination in the $\{111\}$ and $\{11\bar{1}\}$ directions considered in Sec.~\ref{sec:surface}. The crystal can terminate either at the triangular (T) or the kagome (K) lattice, and for $\delta\neq 0$ we further have to specify whether the bonds connecting the outermost layer of sites to the next one are strong (S) or weak (W). Only TW and KS terminations are possible on the top, and only TS and KW are possible on the bottom of a sample in the slab geometry.}
  \label{fig:surf-terminations}
\end{figure}

The SBZ has the shape of a regular hexagon and its construction is indicated in Fig.~\ref{fig:TIorNI}(a). Since the twelve Weyl nodes project onto separate points, six distinct Fermi arcs are expected. Upon increasing the staggered strain $\delta$, the projections of the Weyl nodes move to the inside of the SBZ along the trajectories shown in Fig.~\ref{fig:TIorNI}(b).  A summary of the surface states for the four different terminations shown in Fig.~\ref{fig:surf-terminations} and varying values of $\delta$ [the remaining parameters are fixed as in Eq.~\eqref{eqn:subChosenParam}] is provided in Fig.~\ref{fig:surf-states1}, where we show density plots of the the surface spectral function.

\begin{figure}
  \begin{center}
	\includegraphics[width=0.475\textwidth]{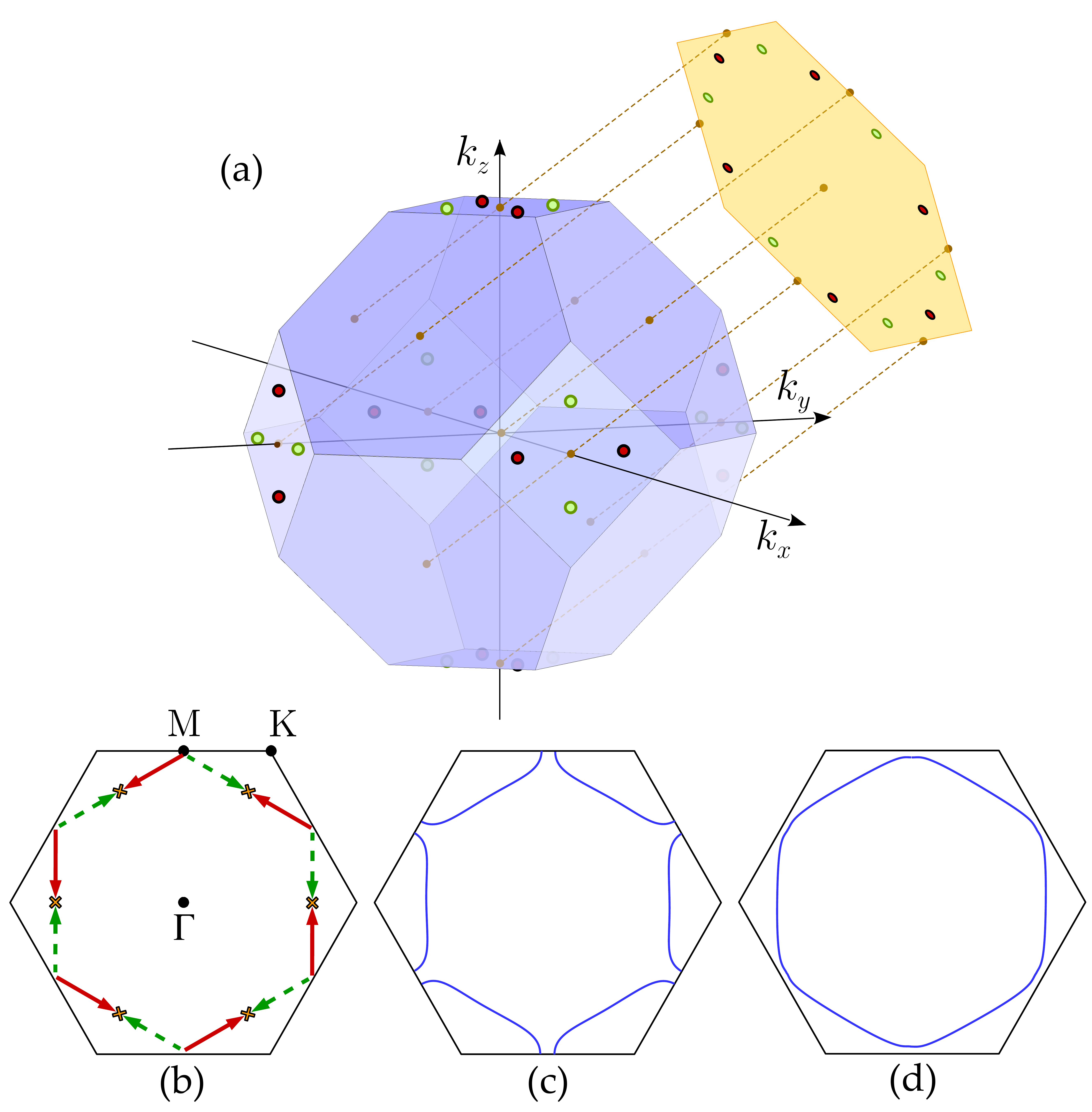}  		
  \end{center}
  \caption{(Color online) (a) Construction of the surface Brillouin zone (SBZ) along the $\{111\}$ directions. The time-reversal invariant momenta (TRIMs) of the SBZ are given by projections of the bulk TRIMs. The projections of the Weyl nodes of the two possible chiralites are also indicated. (b) Definition of the high-symmetry points $\Gamma$, \textrm{M} and $\textrm{K}$ in the SBZ. The three inequivalent $\textrm{M}$ points and the $\Gamma$ point are TRIMs. The solid red and dashed green lines indicate the trajectory of the projections of the Weyl nodes of opposite chiralities upon increasing the value of $\delta$. The crosses indicate where the Weyl points annihilate for the critical $\delta_\textrm{c}$. (c) and (d) The DSM phase gets gapped for finite thickness of the system. The $\mathbb{Z}_2$ invariant of the insulating phase depends on the number of TRIMs encircled by the surface Fermi lines. Situation (c) corresponds to a normal and (d) to a topological insulator.}
  \label{fig:TIorNI}
\end{figure}
\begin{figure*}
  \begin{center}
	\includegraphics[width=0.99\textwidth]{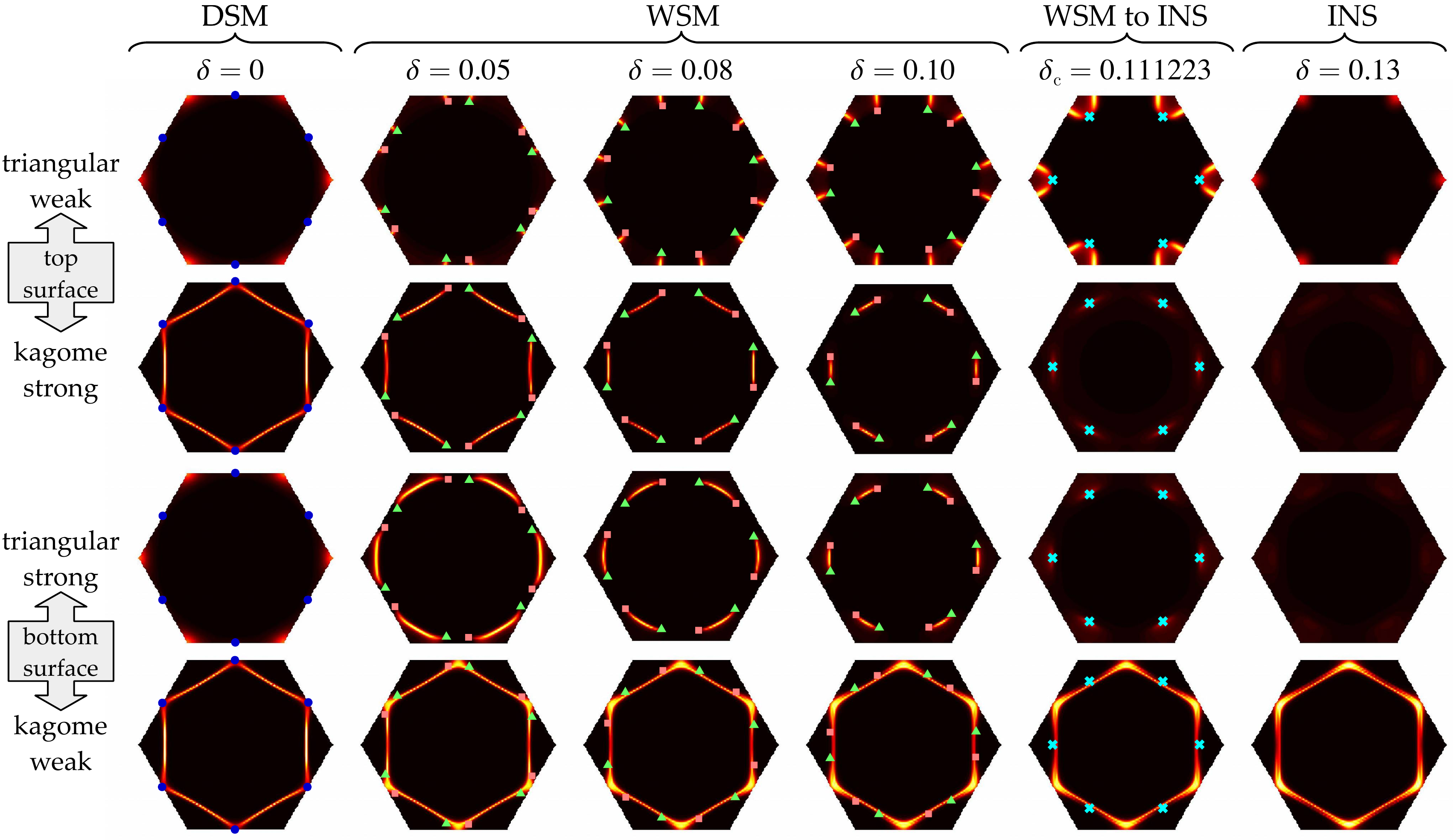}  		
  \end{center}
  \caption{(Color online) Surface spectral function within the SBZ defined in Fig.~\ref{fig:TIorNI}(a), plotted for varying value of $\delta$ and varying surface terminations illustrated in Fig.~\ref{fig:surf-terminations}. The blue circles in the DSM column represent projections of the bulk Dirac nodes into the SBZ. In the same way, the red square and the green triangle markers in the WSM columns represent the Weyl nodes of the two opposite chiralities, and the pale crosses in the WSM-to-INS column represent the $\bs{k}$-points where pairs of Weyl nodes annihilate. In the INS phase, the chemical potential is set to the middle of the bulk gap.}
  \label{fig:surf-states1}
\end{figure*}
\subsection{Strain dependent surface states}
\subsubsection{Surface states of the Dirac semimetal}
The column $\delta = 0$ in Fig.~\ref{fig:surf-states1} corresponds to the DSM phase. We find no surface states for the triangular termination. For the kagome termination, we find a single non-degenerate Fermi line connecting the $\textrm{M}$ points of the hexagonal SBZ, which are the projections of the bulk Dirac nodes.

Since the surface breaks symmetries that protect the bulk Dirac nodes, a small gap, which shrinks with increasing system width,  opens in the (111) film. As a consequence, the surface states for the kagome termination of the DSM phase do not exactly pass through the M points but slightly avoid them. In general, for band insulators, the Fermi line has to encircle an even or odd number of time-reversal invariant momenta (TRIMs) of the SBZ, depending on the strong $\mathbb{Z}_2$ invariant of the band structure \cite{Hasan:2010,Qi:2011}. We therefore expect that either the situation shown in Fig.~\ref{fig:TIorNI}(c) (normal insulator) or in Fig.~\ref{fig:TIorNI}(d) (topological insulator) is realized. We numerically find that the first possibility (normal insulator) is realized, which is consistent with the fact that there are no surface states for the triangular termination.

\subsubsection{Surface states of the Weyl semimetal}
In the WSM phase, the surface states form Fermi arcs connecting the projections of the Weyl nodes. Interestingly, the connectivity of the Fermi arcs, i.e.~the way the Weyl nodes of opposite chirality are paired into the arcs, depends on the termination of the sample. In Ref.~\cite{Hosur:2012}, a similar dependence appears naturally in a toy model consisting of a stack of alternating electron and hole Fermi surfaces. Here, we directly observe this phenomena in a microscopic model. 

For the TW termination, we observe six Fermi arcs developing between the projections of the Weyl points in neighboring SBZs as $\delta$ increases from zero. At the critical $\delta_\textrm{c}$, three Fermi arcs at the time form a closed Fermi line enclosing K and K', respectively. For even larger values of $\delta$, all surface bands are shifted away from the bulk chemical potential. 

For the KS termination, the very opposite happens. The closed Fermi lines present in the DSM phase splits into six distinct Fermi arcs upon breaking the inversion symmetry. Increasing the staggered strain $\delta$ leads to a shrinking of the Fermi arcs and to a complete disappearance of the surface states at $\delta_\textrm{c}$.

For the TS termination and $\delta\gtrsim 0.04$, the same pairs Weyl nodes as for the KS termination are connected but with opposite curvature. They also disappear at $\delta_\textrm{c}$. In Subsec.~\ref{sec:WL}, we separately discuss the interesting parameter range $0<\delta\lesssim 0.04$. 

Finally, for the KW termination, additional Fermi arcs appear very close to the Fermi lines present in the DSM phase. This fact makes their observation obscured in Fig.~\ref{fig:surf-states1}. For the INS phase we find a pair of closed Fermi lines encircling the $\Gamma$ point.

\subsubsection{Surface states of the insulator}
In the INS phase, we observe no surface states for the TS, TW and KS terminations. However, for the KW termination, there are two surface bands crossing the bulk chemical potential that encircle the $\Gamma$ point. These observations are in accordance with the trivial $\mathbb{Z}_2$ invariant of the INS phase.

One possibility to understand the ``non-topological" surface states of the KW termination is to consider the (unphysical) limit $\delta\to 1$ and $p\to 0$ which corresponds to a lattice of \emph{isolated} tetrahedra.  A lattice of isolated tetrahedra exhibits only flat bands. The boundary consists of 
\begin{itemize}
\item[(i)] isolated tetrahedra for the TS and KS terminations,
\item[(ii)] isolated triangles for the KW termination, 
\item[(iii)] isolated points for the TW termination. 
\end{itemize} 
In the situation (i), the bulk and the surface have identical spectra so the chemical potential lies in the gap of both. For sufficiently weak coupling between the tetrahedra, this observation has to remain valid and we do not expect surface states for the TS and KS terminations.

For the other two situations, we need to know the eigenenergies of the corresponding boundary objects. Focusing on $s=+1$, the states of an isolated tetrahedron with spin-orbit coupling parametrized by $R$ lie at energies
\begin{equation}
\epsilon_1^{(4)} = 6,\quad \epsilon_2^{(4)} = -2(1-4R),\quad \epsilon_{3,4}^{(4)} = -2(1+2R).
\end{equation}
Due to time-reversal symmetry, all levels are doubly degenerate. It follows that for $R=-0.4$ (as in Fig.~\ref{fig:surf-states1}), the chemical potential satisfies $\epsilon_{3,4}^{(4)}<\mu<\epsilon_1^{(4)}$. On the other hand, the energy levels of an isolated triangle appearing at the KW termination are
\begin{align}
\epsilon_{\pm}^{(3)} &= 1+2R \pm\sqrt{3\left[3+4(R-1)R\right]},\\
\epsilon_{3}^{(3)} &= -2(1+2R).
\end{align}
Hence, the highest energy level $\epsilon_+^{(3)}$ also satisfies $\epsilon_{3,4}^{(4)}<\epsilon_+^{(3)}<\epsilon_1^{(4)}$ which pins $\mu$ at $\epsilon_+^{(3)}$ for $T=0$ in the charge neutral system. The eigenenergy of an isolated point appearing at the TW termination is simply
\begin{equation}
\epsilon_1^{(1)}=0,
\end{equation}
which too lies between $\epsilon_{3,4}^{(4)}$ and $\epsilon_1^{(4)}$ for $R=-0.4$. We therefore expect that there are surface states in the bulk gap both for the KW and the TW termination, even if we reintroduce the coupling between the tetrahedra. We indeed do observe this. In Fig.~\ref{fig:surf-states1}, we fixed the chemical potential in the INS phase at the bulk value. For the KW termination, the surface states cross the bulk chemical potential, for the TW termination they lie below it. 
\begin{figure}[b!]
  \begin{center}
	\includegraphics[width=0.42\textwidth]{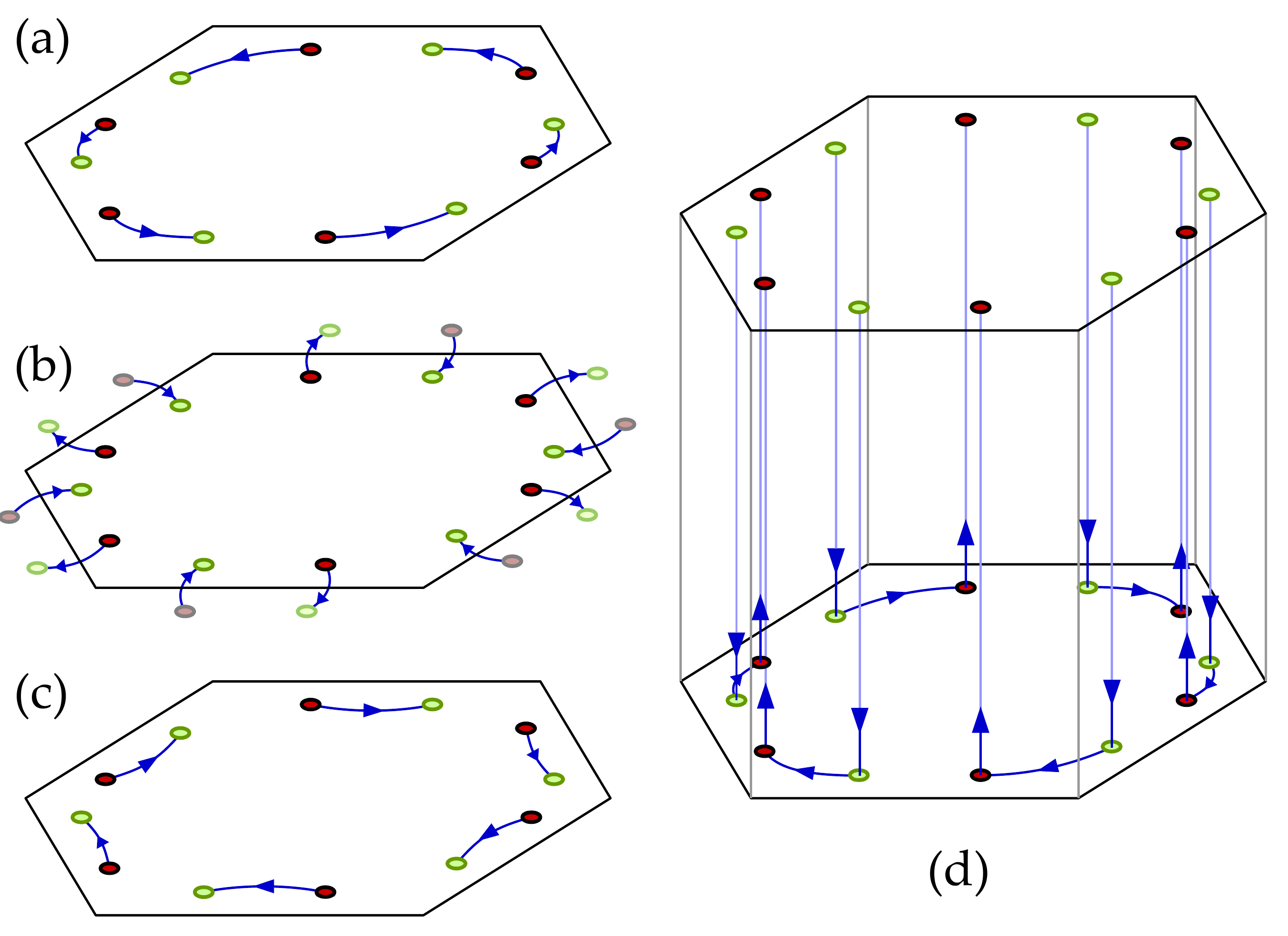}  		
  \end{center}
  \caption{(Color online) (a,b,c) Three connectivities of the Fermi arcs that preserve the symmetry of the system. The dark red and pale green points correspond to the projections of the bulk Weyl nodes of opposite chiralities. The arrows indicate direction of motion of the electrons in the quantum oscillations experiment proposed in~\cite{Potter:2014} when a magnetic field is applied perpendicular to the surface. (d) The oriented vertical lines correspond to the chiral Landau levels emanating from the Weyl nodes, and act as ``conveyor belts" transporting the electrons between the top and bottom surfaces of the slab. If the indicated connectivity of the Fermi arcs is realized on the bottom surface, the electron orbits traverse the bulk 2, 6 and 12 times, respectively, if connectivities (a),(b) and (c) are realized on the top surface. Realizing situations (b) on the top and (c) on the bottom leads to electron orbits traversing the bulk four times.}
  \label{fig:connectivities}
\end{figure}
\subsection{Weyl-Lifshitz transitions}\label{sec:WL}
As illustrated in Fig.~\ref{fig:surf-states1}, the connectivity and shape of the Fermi arcs in the WSM phase depend on the termination of the sample. For the studied surfaces of our model, there are three possible connectivities that respect the symmetries of the system and that do not contain Fermi arcs crossings (which are expected to be gapped out) [Fig.~\ref{fig:connectivities}(a-c)]. Interestingly, it is possible to continuously move from one connectivity to another one, giving rise to a Lifshitz transition of the Fermi arcs that we dub \emph{Weyl-Lifshitz transition}. More precisely, the Weyl-Lifishitz transition is characterized by a topological change of the closed Fermi lines formed by the Fermi arcs of the top and bottom surfaces.

We have observed such a transition for the TS termination for varying staggered strain: For small $\delta$, the connectivity of the Fermi arcs is identical to that of the TW termination, see Fig.~\ref{fig:surf-states1}. At $\delta\approx 0.015$, a part of the topological surface band crosses the chemical potential from below at the two $\textrm{K}$ points. The corresponding Fermi lines grow and at $\delta \approx 0.04$ they touch the original Fermi arcs. A reconnection of the Fermi arcs from situation in Fig.~\ref{fig:connectivities}(b) to that of Fig.~\ref{fig:connectivities}(a) occurs. 

Another possibility is to tune the shape and connectivity of the Fermi arcs by applying a surface gate potential. We model such an experiment in a simplified manner by adding an on-site potential $V$ to the outermost layer of atoms. This effectively shifts the energy of the surface states and those that were originally away from the Fermi level can be tuned to cross it. The surface states are then allowed to hybridize with the Fermi arcs, leading to a reconnection of the Weyl nodes. Figure~\ref{fig:Lifshitz}(a) shows the transition between the connectivity shown in Fig.~\ref{fig:connectivities}(a) and the one shown in Fig.~\ref{fig:connectivities}(b) by applying a positive surface potential to the KS termination.
Similarly, we observe a transition from the situation shown in Fig.~\ref{fig:connectivities}(a) to the one shown in Fig.~\ref{fig:connectivities}(c) if we apply a negative surface potential.
\begin{figure}[t!]
  \begin{center}
	\includegraphics[width=0.49\textwidth]{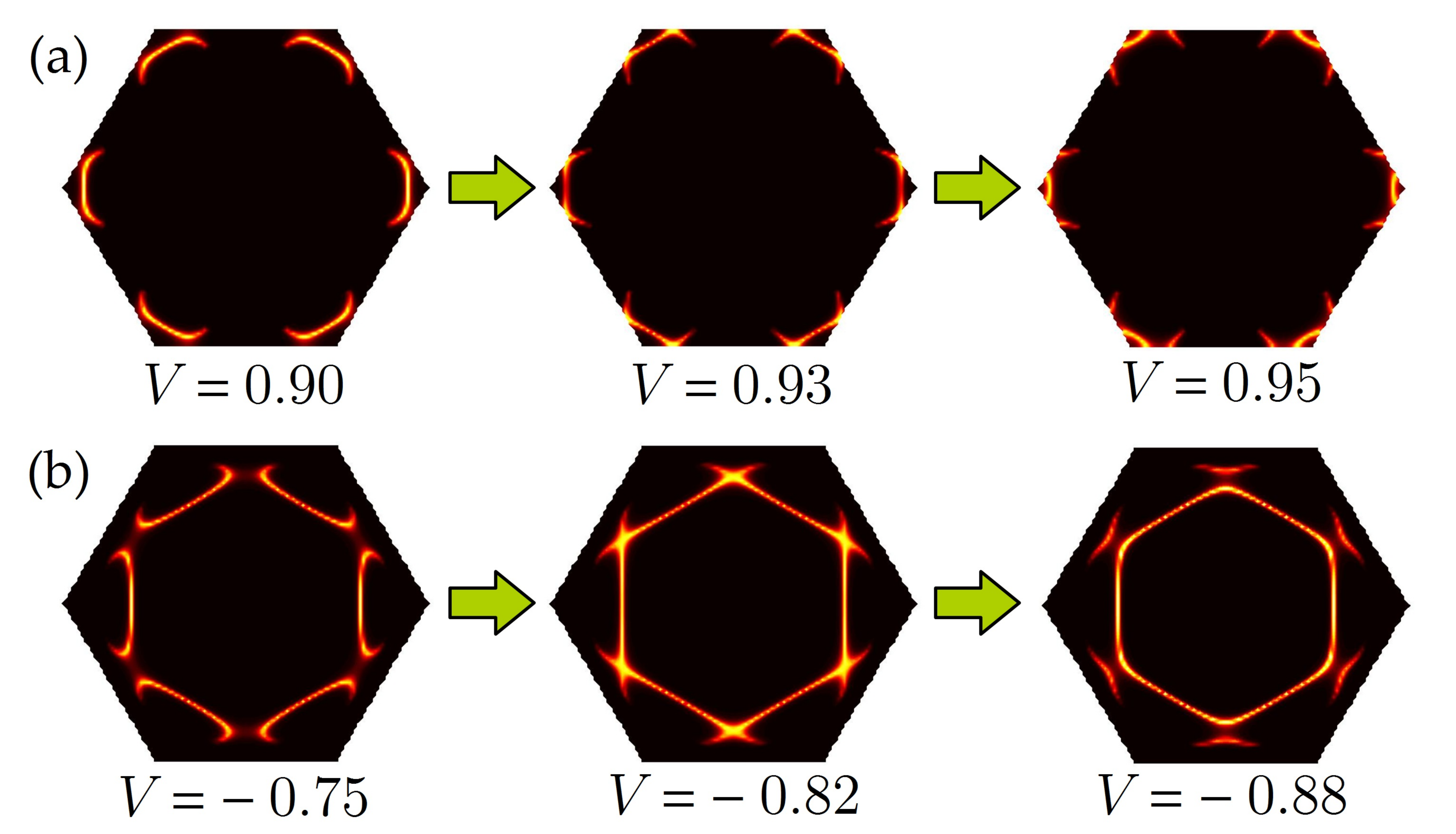}  		
  \end{center}
  \caption{(Color online) Weyl-Lifshitz transitions at $\delta=0.08$ realized by tuning the on-site potential on the outermost layer of atoms for the KS termination. Assuming the TS termination on the opposite surface according to Fig.~\ref{fig:connectivities}(d), we find that (a) the transition at positive $V=0.93$ changes the number of times the electron orbits cross the bulk from 2 to 6, and (b) the transition at negative $V=-0.82$ changes the same number from 2 to 12.}
  \label{fig:Lifshitz}
\end{figure}

A fascinating aspect of the Weyl-Lifshitz transition is that it changes the number of times an electron crosses the bulk in order to complete its semiclassical orbit in an external magnetic field. For example, if we assume the TS termination on the bottom and the KS termination on the top surfaces, the transitions shown in Fig.~\ref{fig:Lifshitz} changes the number of bulk crossings from 2 to 6 in (a) and from 2 to 12 in (b). 

Such a dramatic change might be observable in quantum oscillation experiments in very clean samples for which the mean-free path $l$ exceeds the system width $L$. In fact, according to Ref.~\cite{Potter:2014}, the surface-state response shows periodic-in-$1/B$ oscillations for fixed $\mu$ in the resistivity (Schubnikov-de Hass effect) or magnetization (de Haas-van Alphen effect), which results from energy levels periodically crossing the chemical potential $\mu$. In the semiclassical approximation, this happens for the $n$th energy level if 
\begin{equation}
\frac{1}{B_n}=\frac{e \pi v (n+\gamma)}{k_0\mu}\left|\cos\phi\right|+\frac{ecL}{\hbar k_0}.
\label{eq:QO}
\end{equation}
Here, $k_0$ denotes the Fermi arc length, $v$ the characteristic Fermi velocity, $\phi$ the angle of the magnetic field with respect to the surface normal and $L$ the thickness of the sample. $\gamma$ is of order unity and incorporates small $n$ quantum effects and $2c$ denotes the number of bulk crossings (in Ref.~\cite{Potter:2014}, only $c=1$ has been considered). Hence, by measuring the dependence on the field-direction, one can extract the second term $\frac{ecL}{\hbar k_0}$ of Eq.~\eqref{eq:QO}. In an experiment with a surface gate potential similar to Fig.~\ref{fig:Lifshitz}, the change in $\frac{e c L}{\hbar k_0}$ originates from a change of the length of the Fermi arcs $\Delta k_0$ but also from the change in the number of bulk crossings $\Delta c$. 


\section{Conclusion}\label{sec:conclusion}
In summary, we studied an effective lattice model of spin-orbit coupled electrons on the pyrochlore lattice at a commensurate filling that realizes a Dirac semimetal phase protected by the non-symmorphic space group. Upon coupling the electronic degrees of freedom to the lattice, a staggered strain, which breaks the inversion symmetry, can spontaneously develop. For increasing strain, each Dirac node splits into four Weyl nodes and a topological semimetal phase is realized. At a critical strain, Weyl nodes of opposite chirality annihilate and the system enters a trivial insulating phase. We identified several interesting features of our model such as a reentrant behavior of the Weyl semimetal phase in the elasticity vs.~temperature phase diagram and the appearance of a Griffiths wing in the presence of a staggered stress.

We furthermore presented a detailed group-theoretical analysis of the electronic spectrum, which is independent of the details of the considered lattice model and only relies on symmetries. We highlighted the importance of non-symmorphicity for realizing a Dirac semimetal and displayed symmetry-based arguments for the appearance of the Weyl nodes in the inversion-symmetry broken phase.

Eventually, we studied the surface states of our model. Most notably, we found a dependence of the connectivity and the shape of the Fermi arcs on the termination of the sample.
We also demonstrated that it is possible to continuously change the connectivity of the Fermi arcs through a Weyl-Lifshitz transition by applying a surface potential. A change in the connectivity may also change the number of times an electron crosses the bulk in order to complete its semiclassical orbit, which leads to a clear signal in a quantum oscillation experiment with a varying field direction.

The model considered in this paper is relevant for pyrochlore oxides with heavy transition-metal ions such as iridates. For these systems, one possibility to tune the chemical potential to the Dirac nodes is to consider alloys of the type A$_{2-x}$B$_x$Ir$_2$O$_7$. In transition-metal oxides, electronic correlations are often important but in 5d systems, they are in general less pronounced. Moreover, the fundamental Berry curvature structure around the Weyl nodes is perturbatively stable against interactions \cite{Witczak-Krempa:2014}. In our study, we have therefore neglected these effects and instead studied a non-interacting system. However, the interplay between electron-electron and electron-lattice interactions is a challenging but interesting research direction which we leave for further studies.

In conclusion, we presented a route in which a non-magnetic Weyl semimetal is realized as a thermodynamic phase with spontaneous inversion symmetry breaking upon coupling the electronic degrees of freedom to the lattice. Many of the observed phenomena of our model, such as the reentrant behavior of the Weyl semimetal, the dependence of the surface Fermi arcs on the termination and the Weyl-Lifshitz transitions are expected to be independent of the details of the system and should be applicable to a large class of materials, in particular including other systems belonging to the $\# 227$ ($Fd\bar{3}m$) space group such as diamond, $\beta$-cristobalite and spinel oxides structures.


\begin{acknowledgements}
We would like to thank Adrien Bouhon, Sarah Etter, and Markus Legner for fruitful discussions. We acknowledge financial support through ETH grant 07~13-2 and the Ambizione Program of the Swiss National Science Foundation. 
\end{acknowledgements}

\appendix


\section{Projective representations of space groups}\label{app:groups1}

Here, we provide a derivation of equation (\ref{eqn:projrec}). A more complete discussion can be found in Ref.~\cite{Bradley:1972}. For brevity, we write IR for irreducible representation and $n$DIR for $n$-dimensional IR throughout the appendix.

Every element of a space group $G$ can be expressed as a point operation $R$ followed by a uniform shift by a vector $\bs{t}$ which we write compactly as $\cosrep{R}{t}$,
	\begin{subequations}\label{eqn:sub:trans}
	\begin{equation}
	\cosrep{R}{t}: \bs{v}\mapsto R\bs{v} + \bs{t} \label{eqn:transform3}
	\end{equation}
Action of the element on functions in real space, including Bloch wave functions, is 
	\begin{equation}
	\cosrep{R}{t}: f(\bs{v})\mapsto f[R^{-1}(\bs{v} - \bs{t})]\label{eqn:transform2}
	\end{equation}
and the composition rule is 
	\begin{equation}
	\cosrepnb{R_2}{\bs{v}_2}\circ\cosrepnb{R_1}{\bs{v}_1} = \cosrepnb{R_2 \circ R_1}{R_2\bs{v}_1+\bs{v}_2}.\label{eqn:transform4}
	\end{equation}
	\end{subequations}
The identity element is $\cosrep{E}{0}$.

The set $T$ of all pure translations by Bravais lattice vectors is a subset of $G$, and can be used to write all elements of $G$ compactly as
	\begin{equation}
	G=T\circ\cosrepnb{R_1}{\bs{t}_1}+\ldots+T\circ\cosrepnb{R_n}{\bs{t}_n}\label{eqn:Gcosreps}
	\end{equation}
where the point operations $R_1,\ldots,R_n$ act at the \emph{same point} in real space and are \emph{all different}. If the vectors $\bs{t}_i$ can all be made zero by a proper choice of the point of symmetry, the lattice is called \emph{symmorphic}. If this cannot be done, the lattice is dubbed \emph{non-symmorphic}. Note also that the set 
	\begin{equation}
	F=\left\{R_1,\ldots,R_n\right\}
	\end{equation}
of the point operations is always a group, while the \emph{coset representatives} $\cosrepnb{R_i}{\bs{t}_i}$ form a group if and only if the space group is symmorphic. 

If we take spin-orbit coupling into account, a $2\pi$-rotation of a wave function in real space is accompanied by a $2\pi$-rotation of the electron spin which results in a sign change of the wave function. This operation is \emph{not} equivalent to identity and we denote it as $\overline{E}$. A $4\pi$-rotation is equivalent to identity $E$.

We adopt the periodic boundary conditions. Then the group $T$ is Abelian and as such it has only 1DIRs $\rho_{\bs{k}}$ labelled by momenta $\bs{k}$. The corresponding representation space is spanned by a Bloch wave function at $\bs{k}$
	\begin{equation}
	\psi_{\alpha,\bs{k}}(\bs{r})=\exp{(\mathrm{i}\bs{k}\cdot\bs{r})}u_{\alpha,\bs{k}}(\bs{r})\label{eqn:BlochWF}
	\end{equation}
where $u_{\alpha,\bs{k}}(\bs{r})$ is the cell-periodic part. The representations of a pure translation by vector $\bs{v}$ is 
	\begin{equation}
	\rho_{\bs{k}}(\cosrep{E}{\bs{v}})=\exp{\left(-\mathrm{i}\bs{k}\cdot\bs{v}\right)}.\label{eqn:1DIRofT}
	\end{equation}

The IRs of the space group $G$ can be more than one-dimensional, but they reduce to the 1DIRs (\ref{eqn:1DIRofT}) on the subgroup $T$. Let us consider a representation $\rho$ of $G$ that contains $\rho_{\bs{k}}$ in its decomposition on subgroup $T$, i.e. it contains $\psi_{\alpha,\bs{k}}(\bs{r})$ as one of the basis vectors in its representation space. Then element $\cosrep{R}{t}$ transforms a Bloch function at $\bs{k}$ into 
	\begin{align}
	\cosrep{R}{t}: & \exp{(\mathrm{i}\bs{k}\cdot\bs{r})}u_{\alpha,\bs{k}}(\bs{r}) \mapsto \nonumber \\
	\mapsto & \exp{\left[\mathrm{i}\bs{k}\cdot(R^{-1}\bs{r})-\mathrm{i}\bs{k}\cdot\bs{t}\right]} u_{\alpha,\bs{k}}(R^{-1}\bs{r}-\bs{t}) 
	\end{align}
which can be easily recognized as a Bloch function at $R\bs{k}$. This means that $\rho$ necessarily also contains $\rho_{R\bs{k}}$ in its decomposition on $T$.

To find the allowed spectrum degeneracies at $\bs{k}$ we have to consider only those symmetry operations that leave the momentum of a Bloch function invariant (modulo reciprocal lattice vectors). We construct it as follows. Let us denote the subgroup of point operations $F$ that leave $\bs{k}$ invariant (called the \emph{little co-group of} $\bs{k}$) as $\overline{G}^{\bs{k}}$. Then the group we are looking for is
	\begin{equation}
	G^{\bs{k}}=\bigcup_{i}T\circ\cosrepnb{R_i}{\bs{t}_i},\quad R_i\in \overline{G}^{\bs{k}}.\label{eqn:LGcosrep}
	\end{equation}
It is a subgroup of $G$ called the \emph{little group of} $\bs{k}$. 

The IRs $\widetilde{\rho}_{\bs{k}}$ of $G^{\bs{k}}$ reduce on the subgroup $T$ to 1DIRs labelled by the \emph{same} momentum $\bs{k}$, so according to equation (\ref{eqn:1DIRofT}) for a Bravais vector $\bs{v}$
	\begin{equation}
	\widetilde{\rho}_{\bs{k}}\left(\cosrep{E}{\bs{v}}\right)=\exp{(-\mathrm{i}\bs{k}\cdot\bs{v})}\mathbb{1}\label{eqn:scalrep}
	\end{equation}
where $\mathbb{1}$ is the unit matrix. This means that representations of Bravais translations commute with representations of all other elements of $G^{\bs{k}}$.

It is useful to perform a substitution
	\begin{equation}
	\widetilde{\rho}_{\bs{k}}(\cosrep{R}{t})=\exp{(-\mathrm{i}\bs{k}\cdot\bs{t})} D_{\bs{k}}(\cosrep{R}{t}).\label{eqn:RepSubst}
	\end{equation}
The composition rule (\ref{eqn:transform4}) and the representation of Bravais translations (\ref{eqn:scalrep}) imply that
	\begin{align}
	D_{\bs{k}}(\cosrepnb{R_i}{\bs{t}_i})&D_{\bs{k}}(\cosrepnb{R_j}{\bs{t}_j})\nonumber\\
	=&\exp{(-i\bs{g}_i\cdot\bs{t}_j)}D_{\bs{k}}(\cosrepnb{R_k}{\bs{t}_k})
	\end{align}
where $\bs{g}_i=\left(R_i^{-1}\bs{k}\right) - \bs{k}$ is a reciprocal lattice vector, $R_k=R_i \circ R_j$ is a point group operation from $\overline{G}^{\bs{k}}$, and $\bs{t}_k$ is a vector appearing together with $R_k$ in expansion (\ref{eqn:Gcosreps}). Note that instead of considering the function $D_{\bs{k}}$ on elements of $G^{\bs{k}}$, we might restrict our attention to its values on elements of $\overline{G}^{\bs{k}}$ by defining
	\begin{subequations}\label{eqn:REPonGbar}
	\begin{align}
	\overline{D}_{\bs{k}}(R_i)&:=D_{\bs{k}}(\cosrepnb{R_i}{\bs{t}_i}),\label{eqn:REPonGbar1} \\
	\overline{D}_{\bs{k}}(R_i)\overline{D}_{\bs{k}}(R_j)&= \exp{(-i\bs{g}_i\cdot\bs{t}_j)} \overline{D}_{\bs{k}}(R_i \circ R_j).\label{eqn:projrecAPPEND}
	\end{align}
	\end{subequations}
This completes the derivation of equation (\ref{eqn:projrec}). 

The factors $\exp{(-i\bs{g}_i\cdot\bs{t}_j)}$ are completely fixed by the crystal symmetry and are referred to as the \emph{factor system} of the projective representation. The projective representations of group $\overline{G}^{\bs{k}}$ can be found as ordinary representation of some larger group that we will refer to as the \emph{extension group of} $\overline{G}^{\bs{k}}$. One only has to pick up those representations of the extension group that are compatible with the factor system. Reference~\cite{Bradley:1972} goes through all high symmetry point of all space groups, gives the appropriate extension of every little co-group, provides a complete list of their representations, and picks up those that are compatible with the factor system. 


\section{Linear dispersion around the Dirac node}\label{sec:sym-prod}

In subsection~\ref{subsec:X227} we analysed the dispersion around the four-fold degeneracy at the $\textrm{X}$ point by studying constraints posed on the matrix elements~(\ref{eqn:X-PT}) by the generators of $\overline{G}^{\textrm{X}}$. In this appendix, we briefly introduce an alternative procedure mentioned in Ref.~\cite{Young:2012}. This method is computationally very efficient, but it conceals the role of individual symmetries as well as the specific form of the $\bs{k}\cdot \bs{p}$ expansion~(\ref{eqn:227XHam-afterT}).

According to the Clebsh-Gordan decomposition, the number of times, $a_\sigma$, that the complex conjugate representation $\rho^{\sigma *}$ is contained in the product representation $\rho^\mu \times \rho^\nu$, is given by
	\begin{equation}
	a_\sigma = \frac{1}{\abs{G}}\sum_k \chi^\mu(\mathcal{C}_k)\chi^\nu(\mathcal{C}_k)\chi^{\sigma}(\mathcal{C}_k) N_k\label{eqn:ClebschGordan}
	\end{equation}
where $\mathcal{C}_k$ are the classes of the group $G$ (in our case, the extension group of $\overline{G}^{\textrm{X}}$), $\abs{G}$ is the number of elements of $G$, $N_k$ is the number of elements in $\mathcal{C}_k$, and $\chi$ stands for the characters of the representations. The right-hand side of Eq.~(\ref{eqn:ClebschGordan}) is symmetric under permutations of $(\mu,\nu,\sigma)$, therefore~\cite{Hamermesh:1964} $a_\sigma$ is also the number of times $\rho^{\mu *}$ appears in $\rho^{\nu}\times\rho^{\sigma}$, and the number of times $\rho^{\nu *}$ appears in $\rho^{\mu}\times\rho^{\sigma}$.

We further consider the selection rules. Let $\ket{\psi_i^\mu}$ be $d^\mu$ states that belong to IR $\rho^\mu$, and $\ket{\phi_j^\nu}$ be $d^\nu$ states that belong to IR $\rho^\nu$. The scalar products $\braket{\psi_i^\mu}{\phi_j^\nu}$ are invariant under $G$. Using the Schur orthogonality relation, we find
	\begin{align}
	\braket{\psi_i^\mu}{\phi_j^\nu} 
		&= \frac{1}{\abs{G}} \sum_{k\ell} \sum_{g\in G} \rho_{ik}^{* \mu}(g) \rho_{j\ell}^\nu(g) \braket{\psi_k^\mu}{\phi_\ell^\nu} \nonumber\\
		&= \frac{1}{\abs{G}} \sum_{k\ell} \braket{\psi_k^\mu}{\phi_\ell^\nu} \frac{\abs{G}}{d^\mu} \delta_{ij}\delta_{k\ell}\delta^{\mu\nu} \nonumber\\
		&= \frac{1}{d^\mu} \sum_k \braket{\psi_k^\mu}{\phi_k^\nu}\delta_{ij}\delta^{\mu\nu},\label{eqn:SelRules}
	\end{align}
i.e. the products are non-zero iff the functions $\ket{\psi_i^\mu}$ and $\ket{\phi_j^\nu}$ belong to the \emph{same representation}. By the symmetry of (\ref{eqn:ClebschGordan}), this is equivalent to stating that the trivial representation $\rho^1$ is contained in the product representation $\rho^\mu\times\rho^\nu$.

We further consider matrix elements 
	\begin{equation}
	\bra{\psi_i^\mu}\hat{\mathcal{O}}\ket{\phi_j^\nu}\label{eqn:GenMatElem}
	\end{equation}
of an operator $\mathcal{O}$ that transforms according to representation $\rho^\sigma$. It is useful to define states $\big|\tilde{\phi}_k^{(\sigma\times\nu)}\big\rangle = \mathcal{O}_{kj}\ket{\phi_j^\nu}$ that transform according to $\rho^\sigma\times\rho^\nu$. The findings of the previous paragraph applied to states $\ket{\psi_i^\mu}$ and $\big|\tilde{\phi}_k^{(\sigma\times\nu)}\big\rangle$ imply that matrix elements (\ref{eqn:GenMatElem}) are non-zero iff $\rho^\sigma$ is contained in the product $\rho^\mu\times\rho^\nu$~\cite{Hamermesh:1964}.

We apply these findings to matrix elements~(\ref{eqn:X-PT}). As discussed therein, the perturbation Hamiltonian transforms according to the vector representation $\rho^{\textrm{vec}}$, and the states transform according to representation $\rho^{4\textrm{D}}$. Note that we can decompose the product $\rho^{4\textrm{D}}\times\rho^{4\textrm{D}}$ as a sum of the symmetric part $\left[\rho^{4\textrm{D}}\times\rho^{4\textrm{D}}\right]$ acting in the hermitian sector of the matrix elements,
	\begin{equation}
	S_{ij}(\bs{p}) = \left[\mathcal{H}_{\textrm{pert.}}^{\textrm{X}}(\bs{p})\right]_{ij} + \left[\mathcal{H}_{\textrm{pert.}}^{\textrm{X}}(\bs{p})\right]_{ji}^*,
	\end{equation}
and the antisymmetric part $\left\{\rho^{4\textrm{D}} \times \rho^{4\textrm{D}} \right\}$ acting in the antihermitian sector of the matrix elements,
	\begin{equation}
	A_{ij}(\bs{p}) = \left[\mathcal{H}_{\textrm{pert.}}^{\textrm{X}}(\bs{p})\right]_{ij} - \left[\mathcal{H}_{\textrm{pert.}}^{\textrm{X}}(\bs{p})\right]_{ji}^*.
	\end{equation}
Since, in our case, the two sets of wave functions are \emph{identical}, the antisymmetric sector completely vanishes. Hence, the linear correction to the spectrum is non-zero iff $\rho^{\textrm{vec}}$ is contained in the \emph{symmetrized Kronecker product} $\left[\rho^{4\textrm{D}}\times \rho^{4\textrm{D}}\right]$.


\bibliography{bibliography}{}
\bibliographystyle{apsrev4-1}  
\end{document}